\documentclass{aa}  
\usepackage{graphicx}
\usepackage{txfonts}
\usepackage{lscape}
\newcommand{\marc}{$\mathrm{mag~arcsec^{-2}}$}
\begin{document}
   \title{Diffuse stellar emission in X-ray luminous galaxy clusters at z $\sim$ 0.3}

   \subtitle{I. Is the diffuse optical light boosted and rejuvenated in merging clusters?}

   \author{D. Pierini
          \inst{1},
          S. Zibetti
	  \inst{2},
          F. Braglia
          \inst{1},
          H. B\"ohringer
          \inst{1},
          A. Finoguenov,
          \inst{1},
          P. D. Lynam
          \inst{3},
          \and
          Y.-Y. Zhang
          \inst{4}
          }

%   \offprints{D. Pierini}

   \institute{Max-Planck-Institut f\"ur extraterrestrische Physik,
              Giessenbachstrasse, D-85748 Garching bei M\"unchen, Germany\\
              \email{dpierini,fbraglia,hxb,alexis@mpe.mpg.de}
         \and
             Max-Planck-Institut f\"ur Astronomie,
             K\"onigstuhl 17, D-69117 Heidelberg, Germany\\
             \email{zibetti@mpia-hd.mpg.de}
         \and
             ESO,
             Karl-Schwarzschild-Strasse 2, D-85748 Garching bei M\"unchen, Germany\\
             \email{plynam@eso.org}
         \and
             Argelander-Institut f\"ur Astronomie, Rheinische Friedrich-Wilhelms-Universit\"at Bonn,
             Auf dem H\"ugel 71, D-53121 Bonn, Germany\\
             \email{yyzhang@astro.uni-bonn.de}
%             \thanks{}
             }

   \date{Received ..., 2008; accepted ...}

% \abstract{}{}{}{}{} 
% 5 {} token are mandatory
 
\abstract
{Clusters of galaxies host a diffuse population of intergalactic stars.
  Diffuse optical light is observed in clusters up to redshift $z \sim 0.4$.
  Recent cosmological hydrodynamical simulations show that
  this intracluster light originates nearly in parallel
  with the build-up of the brightest cluster galaxy (BCG),
  as identified at $z=0$.
  However, theory proposes alternative scenarios for its origin.}
{We searched for diffuse stellar emission around BCGs
  in three of the most X-ray luminous clusters found at $z \sim 0.3$
  in the REFLEX cluster survey and observed with XMM-\emph{Newton\/}.
  These systems (RXCJ\,0014.3$-$3022, RXCJ\,0232.2$-$4420,
  and RXCJ\,2308.3$-$0211) are in different dynamical states,
  as witnessed by their X-ray morphology and optical appearence
  (e.g. multiplicity of BCGs).}
{Existing medium--deep, wide-field imaging in B and R bands
  allows extension, intensity, and colour of the stellar emission
  to be determined across a region that encompasses
  the X-ray emission from the intracluster medium (ICM) in each cluster.}
{Diffuse stellar emission is robustly detected
  down to a surface brightness of 26~R-\marc~(observed frame)
  around a total of seven BCGs, extending up to galactocentric distances
  of $\sim 100~h_{70}^{-1}~\mathrm{kpc}$.
  In particular, it surrounds a pair of BCGs in RXCJ\,0232.2$-$4420,
  while it bridges two BCGs associated with the minor subcomponent
  of the merging cluster RXCJ\,0014.3$-$3022.
  The diffuse light detected at the greatest distances from the BCGs
  of the rather regular clusters RXCJ\,0232.2$-$4420 and RXCJ\,2308.3$-$0211
  follows the ICM distribution.
  Its $\mathrm{B-R}$ colour is consistent
  with the colours measured within the BCG effective radii.
  The diffuse light around the two pairs of BCGs in RXCJ\,0014.3$-$3022
  exhibits bluer colours than the BCG central regions by up to 0.5 mag.}
{If the contribution of the intracluster light (ICL)
  to the detected diffuse light around BCGs is not negligible, ICL and BCGs
  have similar stellar populations in relatively relaxed clusters.
  Merging on a cluster scale eventually adds gravitational stresses
  to BCGs and other galaxies in subcluster cores.
  This event may affect the properties of the diffuse stellar emission
  around BCGs.
  Shredding of star-forming, low-metallicity dwarf galaxies
  is favoured as the cause of the bluer $\mathrm{B-R}$ colours
  of the diffuse stellar component around the two pairs of BCGs
  in the merging cluster RXCJ\,0014.3$-$3022.}


\titlerunning{Diffuse stellar emission in X-ray luminous galaxy
  clusters at z $\sim$ 0.3}
\authorrunning{Pierini D. et al.}
\maketitle
%
%________________________________________________________________

\section{Introduction}\label{intro}

A diffuse population of intergalactic stars exists in galaxy clusters,
which has been well-established observationally in \emph{individual\/} systems
as well as in a \emph{statistical\/} sense (see Zibetti 2007 for a review).
Observations of this intracluster light (ICL) in individual clusters
now extend up to redshifts $0.3 \la z \la 0.4$ (Krick \& Bernstein 2007;
Rines et al. 2007).
For nearby clusters, the \emph{photometric\/} detection of the ICL
reaches down to very faint surface brightness levels,
from $\sim 26$ down to $\sim 30$ R-\marc~(Bernstein et al. 1995;
Gonzales et al. 2000, 2005; Feldmeier et al. 2002, 2004).
Only in a few such systems, however, the presence of intergalactic stars
\emph{not\/} bound to any galaxy could be convincingly proved
through a \emph{dynamical\/} evidence (Arnaboldi et al. 2004).
In a complementary approach (Zibetti et al. 2005),
stacking images of 683 clusters at $0.2 \la z \la 0.3$
allowed an average surface brightness of the ICL
between $\sim 26$ and $\sim 32$ R-\marc to be reached.
This technique made a photometric detection of the ICL possible 
at much deeper levels, enabling the spatial distribution
and colour of the ICL to be studied at extremely large distances
(up to $700~h_{70}^{-1}~\mathrm{kpc}$) from the brightest cluster galaxy (BCG).
The ICL contributes on average 30--40\% to the total optical emission
(including galaxies) at $\sim 100~h_{70}^{-1}~\mathrm{kpc}$;
this fraction decreases to $\la 5$\% at a galactocentric distance
of 600--$700~h_{70}^{-1}~\mathrm{kpc}$.

On one hand, the radial behaviour of the ICL fraction
and the consistency between the optical colours of ICL and BCGs
point to an origin of the ICL from galaxies
that experience very strong tidal interactions
close to the bottom of a cluster's potential well,
and eventually merge into the brightest cluster galaxy (Zibetti et al. 2005).
This interpretation is consistent with the kinematic evidence that
intracluster planetary nebulae originate from the ongoing subcluster merger
in the Coma cluster core (Gerhard et al. 2007).
However, nearby clusters like Virgo exhibit a wealth of tidal structures
(e.g. Mihos et al. 2005), which witnesses that stars
can be injected all over the cluster environment.
The same conclusion is reached in the study of flux, profile, colour,
and substructure of the ICL in 10 clusters with different redshifts,
X-ray and optical properties (Krick \& Bernstein 2007).
Consistently, the extent of the ICL and its similar elongation
to the cluster galaxy distribution in 24 galaxy clusters at $0.03 < z < 0.13$
suggest that the evolution of the ICL is tied to the cluster as a whole
rather than to the BCG (Gonzalez et al. 2005).
However, this correspondence does not always hold (see Feldmeier et al. 2004).

As for the ICL origin, several mechanisms may be at work to some extent,
as theory proposes (Sommer-Larsen et al. 2005; Rudick et al. 2006;
Murante et al. 2007; Purcell et al. 2007; Conroy et al. 2007).
The bulk of the intracluster stars may originate in (proto)galaxies,
which have been partly or fully disrupted through tidal stripping
in the main cluster potential or by galaxy-galaxy interactions
(Sommer-Larsen et al. 2005).
Conversely, Murante et al. (2007) show that
the formation of the diffuse stellar component in clusters
is parallel to the build-up of the brightest cluster galaxy
and other massive galaxies,
whereas dissolved galaxies contribute only 25\% to the ICL budget.
Finally, for Purcell et al. (2007), different shredded galaxies
are the source of diffuse intrahalo light on varying scales.
The diffuse light in group and cluster halos is built from satellite galaxies
that form stars efficiently;
intracluster light is dominated by material liberated from galaxies
with a stellar mass larger than $\sim 10^{11}~\mathrm{M_{\sun}}$.

A significant fraction, $\sim 30$\%, of stars needs to be scattered
to the diffuse stellar component of clusters at each galaxy-galaxy merging
to account for the little evolution of the high end
of the stellar mass function of galaxies since $z \sim 1$,
which turns to be consistent with the total amount of the ICL
observed in clusters (Monaco et al. 2006).
On the other hand, model brightest cluster galaxies assemble surprisingly late:
half of their final mass is typically locked up in a single galaxy
after $z \sim 0.5$ (De Lucia \& Blaizot 2007).
These two theoretical results prompted our search
for diffuse stellar emission around BCGs
in the 13 most X-ray luminous clusters at $z \sim 0.3$
in the ROSAT-ESO Flux-Limited X-ray (REFLEX) cluster survey
(B\"ohringer et al. 2001) and observed with XMM-\emph{Newton\/}
(Zhang et al. 2006).

Krick \& Bernstein (2007) have already detected diffuse light
in one of the clusters of our sample (i.e., RXCJ\,0014.3$-$3022 alias AC\,118,
see Zhang et al. 2004).
We confirm their discovery and discuss the properties of the diffuse light
around the two pairs of BCGs in this merging system in much greater detail.

Throughout this paper we adopt a $\Lambda$ cold dark matter (CDM)
cosmological model where $\Omega_{\mathrm{m}} = 0.3$,
$\Omega_{\mathrm{\Lambda}} = 0.7$,
and $H_0 = 70~h_{70}~\mathrm{km~s^{-1}~Mpc^{-1}}$,
which is consistent with the main results of Spergel et al. (2007).
Hence a metric separation of 1 Mpc corresponds to an angular distance
of 3.75 arcmin at $z = 0.3$.
%
%__________________________________________________________________

\section{The sample}\label{sample}

For this pilot study, we selected three objects
from the REFLEX Distant X-ray Luminous (DXL) cluster sample
(Zhang et al. 2006) with existing optical imaging.
Table~\ref{XrayProp} lists denomination, redshift, characteristic radius
$R_{500}$\footnote{$R_{500}$ is the radius within which the average density
of the DM in the cluster is 500 times the critical density of the Universe.},
X-ray bolometric luminosity and classification of each cluster
(from Zhang et al. 2006).

In spite of its small size, our sample spans the range in X-ray morphology
(or the dynamical state of a cluster, see Jones \& Forman 1992)
of the most X-ray luminous clusters at $z \sim 0.3$.
It complements the sample of Krick \& Bernstein (2007),
which spans a range in cluster characteristics,
namely redshift ($0.05 \la z \la 0.3$), morphology (i.e., number of BCGs),
spatial projected density (richness class 0--3), and X-ray luminosity
($1.9 \times 10^{44}~\mathrm{erg\, s^{-1}} < L_{\mathrm{X}} < 22 \times 10^{44}~\mathrm{erg\, s^{-1}}$ for the cosmology adopted here).
As the latter sample, ours is X-ray selected
whereas that of Zibetti et al. (2005) was originally selected in the optical
with a maxBGC method (Koester et al. 2007a, 2007b).
%
%__________________________________________________________________

\section{Observations}\label{obs}

The selected objects were observed with XMM-\emph{Newton\/} in AO-1 and AO-3
as part of the REFLEX-DXL cluster sample (Zhang et al. 2006).
All pointed observations were performed with thin filter
for the three detectors of the European Photon Imaging Camera (EPIC)
on XMM-\emph{Newton\/} (Str\"uder et al. 2001; Turner et al. 2001).
The MOS data were taken in Full Frame (FF) mode,
whereas the \emph{pn\/} data were taken in Extended Full Frame (EFF) mode
in AO-1 and FF mode in AO-3.
For \emph{pn\/}, the fractions of the out-of-time (OOT) effect
are 2.32\% and 6.30\% for the EFF mode and FF mode, respectively.
An OOT event file is created and used to statistically remove the OOT effect.
Total exposure times were equal to 18.3 ks, 13.6 ks, and 11.9 ks
for RXCJ\,0014.3$-$3022, RXCJ\,0232.2$-$4420, and RXCJ\,2308.3$-$0211,
respectively.

The same clusters were observed with the Wide Field Imager
(WFI, Baade et al. 1999) mounted at the Cassegrain focus
of the ESO/MPG 2.2m telescope in La Silla, Chile
(Pierini et al., in preparation).
This is a $4 \times 2$ mosaic detector of $\mathrm{2k \times 4k}$ CCDs
with pixel scale of 0.238~\arcsec/pixel, which gives a field of view
per exposure of $34^{\prime} \times 33^{\prime}$.
Data were taken in the B ($\lambda_0 = 4562.52$ \AA),
V ($\lambda_0 = 5395.62$ \AA),
and R$_\mathrm{c}$ ($\lambda_0 = 6517.25$ \AA, hereafter simply R) bands.
In particular the B and R broad-band filters nicely probe
the spectral region across the 4000 \AA~break, which sets some constraints
on the characteristics of the bulk of stellar populations
existing at $z \sim 0.3$ (see Fig.~\ref{FigBC03Mod}).
So only B and R imaging is used in this study.
Observations were performed on September 27--30, 2000
under photometric conditions and with a seeing of 1.0--1.3\arcsec (FWHM).
They were not aimed at detecting diffuse light
but at characterising clusters (see B\"ohringer et al. 2006).
Total exposure times were equal to 4800 s (B), 3600 s (V), and 3600 s (R),
i.e., 6 times shorter than the average total exposure times in V and r bands
of the data collected by Krick \& Bernstein (2007)
at the 2.5m du Pont telescope for their two galaxy clusters at $z = 0.3$
(i.e., AC\,114 and AC\,118).
Twilight-sky flats were taken in the evening and following morning
for each night of observations.
A dithering pattern with dithers up to a few arcmin
was applied between individual exposures (eight per filter).
This reduces large-scale flat-fielding fluctuations on combination.
However, our WFI data prevent a detection of diffuse light
down to the surface brightness levels reached by Krick \& Bernstein (2007)
through their optimized observing strategy.
%
%__________________________________________________________________

\section{Analysis}\label{datared}

\begin{landscape}
\begin{figure}
  \centering
  \includegraphics[width=24cm]{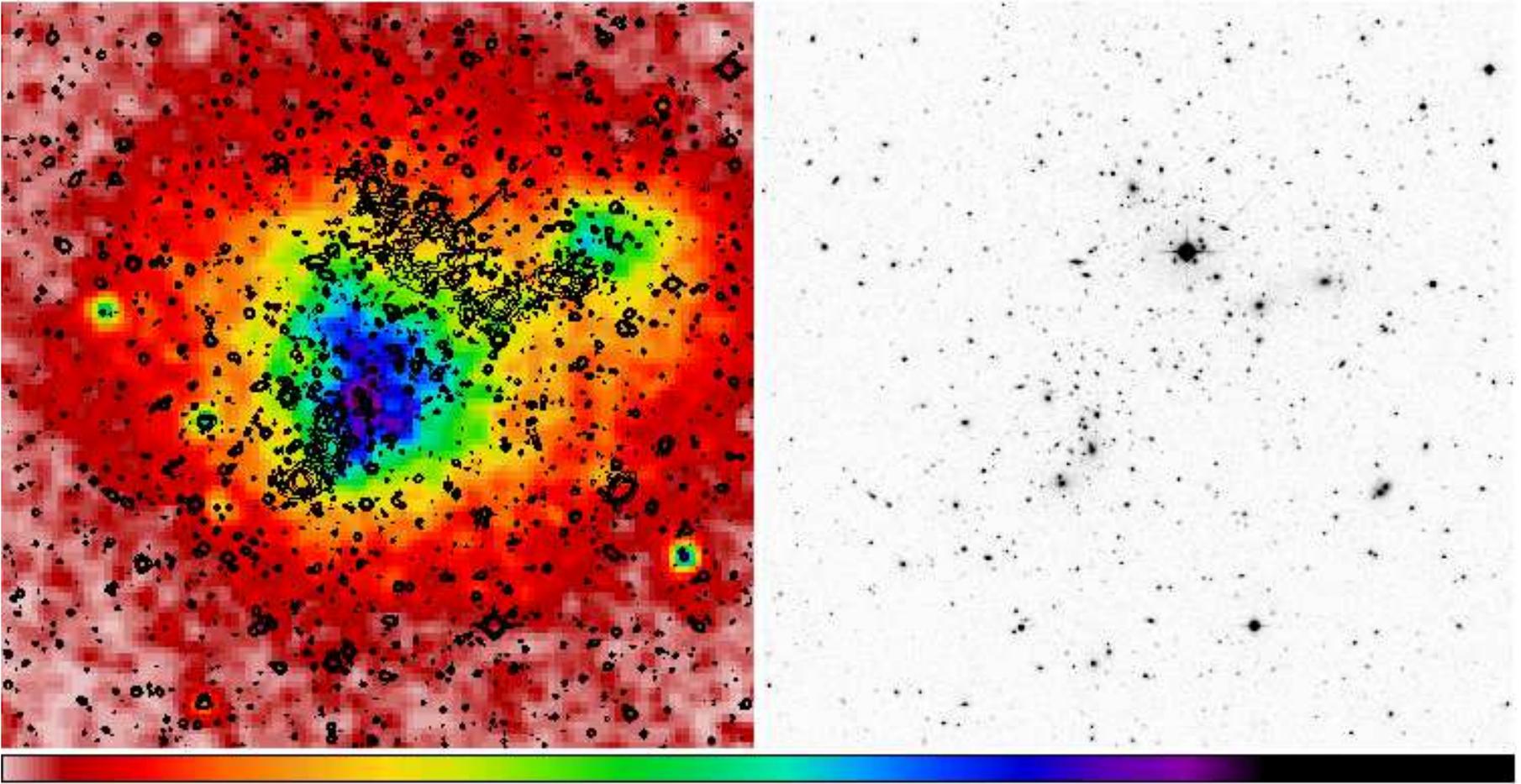}
  \caption{RXCJ\,0014.3$-$3022 as imaged by EPIC (left panel)
    and WFI (right panel).
    A square region, 2.0 Mpc on a side, centred on the X-ray centroid
    of the cluster is reproduced (north is up and east to the left).
    The X-ray surface brightness distribution is reproduced in a map
    using a linear scale where intensity increases as the colour moves
    from the left to the right along the palette at the bottom.
    An overlay with a black contour map reproduces features in the R-band image
    at a significance of 1$\sigma$, 2$\sigma$, 3$\sigma$, 5$\sigma$,
    and 10$\sigma$, the limiting surface brightness being equal
    to 25.74 R-\marc.
    The R-band image to the right (in greyscale)
    highlights high surface brightness regions with a linear scale
    where intensity increases as the tone turns from light grey to black.}
  \label{FigXrayOpt1}
\end{figure}
\end{landscape}

\begin{landscape}
\begin{figure}
  \centering
  \includegraphics[width=24cm]{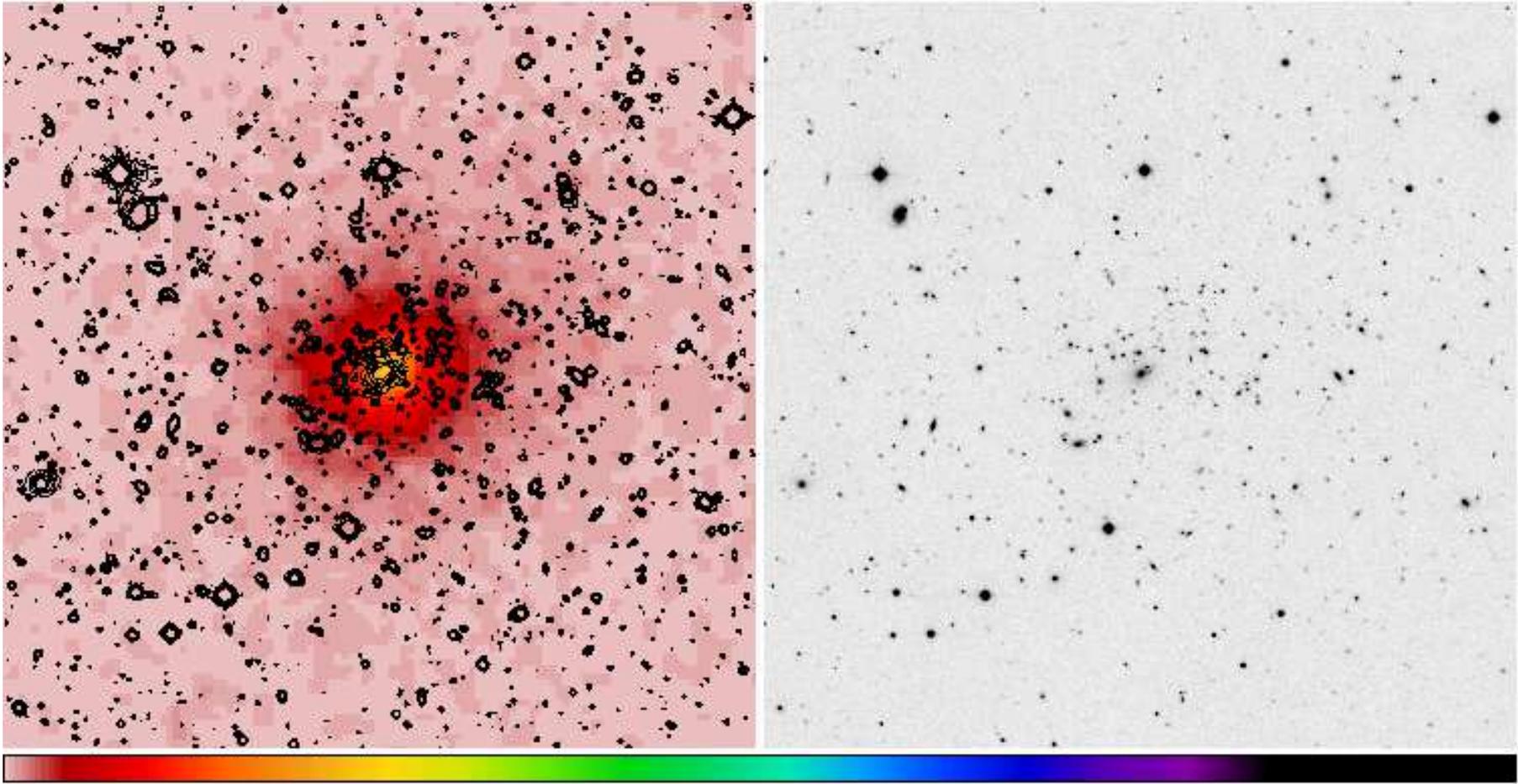}
  \caption{RXCJ\,2308.3$-$0211 as imaged by EPIC and WFI.
    Size and symbols are the same as in Fig.~\ref{FigXrayOpt1}.
    The limiting R-band surface brightness is equal to 25.38~\marc.}
  \label{FigXrayOpt3}
\end{figure}
\end{landscape}

Figs.~\ref{FigXrayOpt1}--\ref{FigXrayOpt3}
reproduce the X-ray surface brighness distribution in a square region
of 2.0 Mpc on a side next to the corresponding R-band image
for the individual clusters.
The X-ray map represents the projected distribution of gas
at temperatures of $10^{7}$--$10^{8}~\mathrm{K}$,
which makes the bulk of the intracluster medium (ICM).
A total of seven BCGs can be clearly identified on the R-band images,
whose basic properties are listed in Table~\ref{BCGProp}.

Details of the EPIC X-ray data reduction, analysis, and results
are described in Zhang et al. (2006).
For the WFI optical data, we made use of the data reduction system
developed for the ESO Imaging Survey (EIS, Renzini \& da Costa 1997)
and its associated EIS/MVM image processing library version 1.0.1
(\emph{Alambic\/}, Vandame 2004).
The B- and R-band combined images of each cluster
were registered and matched to the same seeing.
Magnitudes were calibrated to the Johnson--Cousins filter system
using ancillary observations of standard star fields (Landolt 1992).
They were corrected for atmospheric extinction and galactic extinction
(Schlegel et al. 1998).
They are expressed in the Vega system hereafter.
The photometric quality reached by the optical images is such that
the background RMS on the pixel scale of the R-band images (the deepest ones)
corresponds to 25.74, 25.81, and 25.38~R-\marc~for RXCJ\,0014.3$-$3022,
RXCJ\,0232.2$-$4420, and RXCJ\,2308.3$-$0211, respectively.
For the same images, the 1$\sigma$ background fluctuations amount,
respectively, to 28.25, 27.51, and 27.15~R-\marc,
as measured in 100-pixels wide annuli with inner radius
corresponding to $R_{500}$ centred on each BCG.
Scales of 100 pixels are relevant to the measurement of the diffuse light.
%
%__________________________________________________________________

\subsection{Notes on individual clusters}\label{notes}

\subsubsection{RXCJ\,0014.3$-$3022}

The complex X-ray aspect and galaxy distribution
witness an ongoing cluster merging.
Two visual pairs of BCGs, $\sim 740~\mathrm{kpc}$ apart (projected distance),
can be easily identified in Fig.~\ref{FigXrayOpt1}.
The photometric redshifts of the individual BCGs
are consistent with those of their parent substructures
(from Braglia et al. 2007).
The two X-ray subcomponents of the system are displaced
from their corresponding pairs of BCGs by $\sim 210~\mathrm{kpc}$.
Comparison of X-ray aspect and galaxy distribution
between RXCJ\,0014.3$-$3022 and the famous ``Bullet Cluster''
(see Clowe et al. 2006)
suggests that the former is at an earlier stage of merging,
when the less massive DM subcomponent is still falling in
towards the most massive one.
This is consistent with the conclusion from the substructure analysis
of RXCJ\,0014.3$-$3022 by Braglia et al. (2007).

One visual pair of BCGs lies to the north-west
with respect to the X-ray centroid
(i.e., RXCJ\,0014.3$-$3022\,NW\,A and B).
The projected separation of the two BCGs is $\sim 185~\mathrm{kpc}$.
Lack of spectroscopy for RXCJ\,0014.3$-$3022\,NW\,B
does not permit the physical nature of the pair to be established,
but RXCJ\,0014.3$-$3022\,NW is associated with an X-ray feature
connecting the two subcomponents of the ICM.
At the angular resolution of XMM-\emph{Newton\/} (6~\arcsec~FWHM),
there is no evidence that this X-ray bridge results from the superposition
of two weak point-like sources (i.e., X-ray emitting active galactic nuclei).
The conclusion that this extended X-ray structure
is made of hot gas still confined in the potential well
of RXCJ\,0014.3$-$3022\,NW, in spite of the ongoing large-scale merging,
is tantalizing\footnote{For recent N-body/SPH simulations
of interacting clusters see Mastropietro \& Burkert (2007).}.
The second visual pair of BCGs, to the south-east,
(i.e., RXCJ\,0014.3$-$3022\,SE\,A and B)
is actually made of two very close pairs of galaxies
at consistent photometric redshifts.
A main component (i.e., one of the two BCGs) can be clearly identified
in each of these very close pairs (see also Fig.~\ref{FigBR1b}).
The two BCGs are separated by $\sim 120~\mathrm{kpc}$ on the plane of the sky
and by $4119 \pm 416~\mathrm{km~s^{-1}}$ (rest frame)
along the line of sight (from Braglia et al. 2007).
Hence, likely RXCJ\,0014.3$-$3022\,SE is not an interacting pair.
Its environment can be thought as analogous to the inner region
of the richness 2 cluster A\,3888 at $z=0.15$ (Krick et al. 2006).
%
%__________________________________________________________________

\subsubsection{RXCJ\,0232.2$-$4420}

The distribution of the X-ray emission is not centrally symmetric
(note the broad extention to the south-west in Fig.~\ref{FigXrayOpt2}),
meaning that this cluster is not fully relaxed.
A small, secondary, extended X-ray source is found in the ESE direction.
It is apparently disconnected from the primary X-ray emission component
(i.e., RXCJ\,0232.2$-$4420) at the sensitivity limit
reached by the XMM-\emph{Newton\/} observations.
In the cluster core sits a visual pair of BCGs,
separated by $\sim 100~\mathrm{kpc}$ on the plane of the sky.
One (i.e., RXCJ\,0232.2$-$4420\,A, see Table~\ref{BCGProp})
coincides with the X-ray peak and centroid.
%
%__________________________________________________________________

\subsubsection{RXCJ\,2308.3$-$0211}

The distribution of the X-ray emission is centrally symmetric.
A single BCG is at the centre of this fairly regular cluster.
%
%__________________________________________________________________

\subsection{Detection of a diffuse stellar component}\label{detect}

\begin{figure*}
  \centering
  \includegraphics[width=15cm]{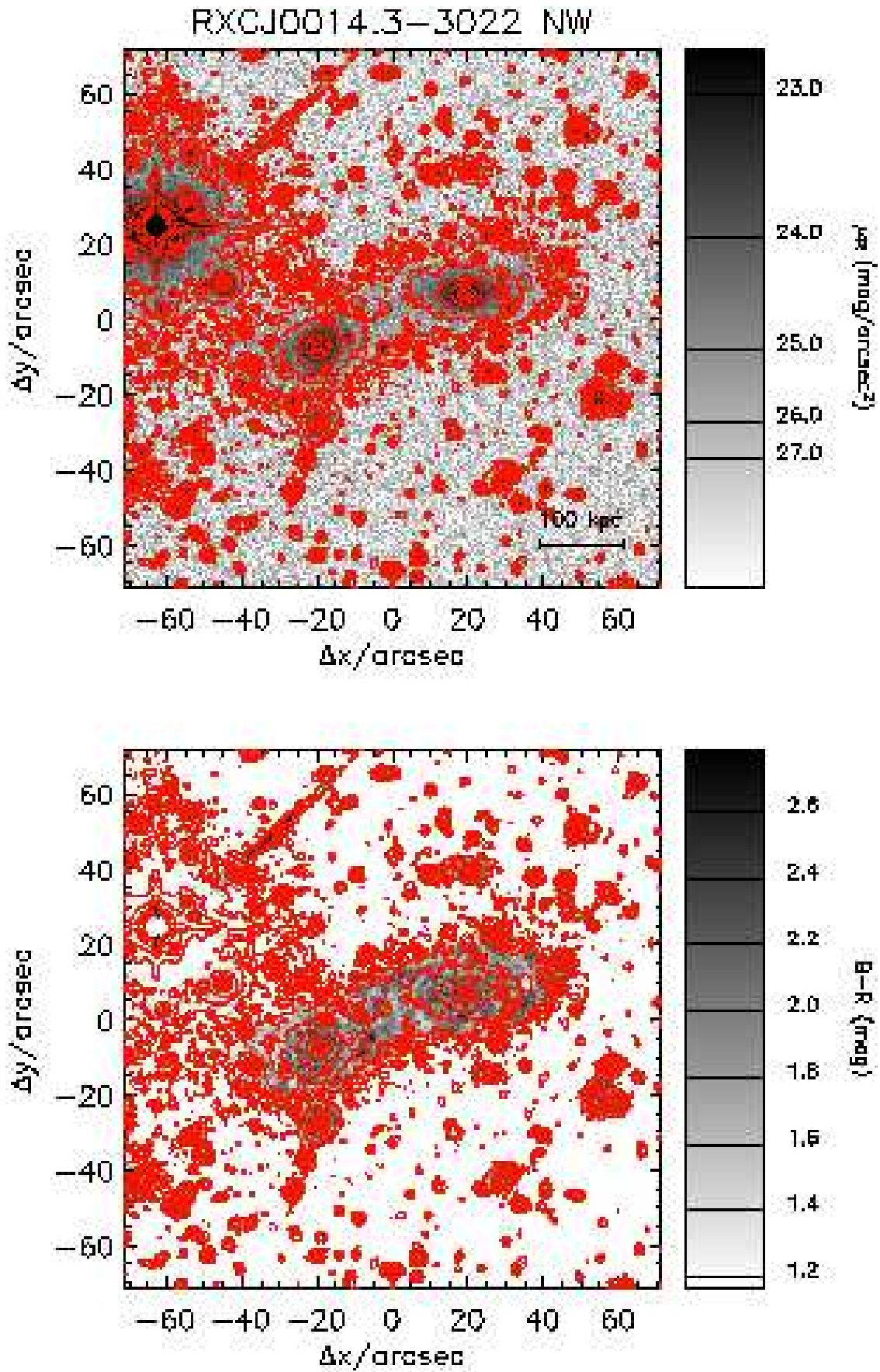}
  \caption{R-band image (top) and adaptively median-smoothed
    $\mathrm{B-R}$ colour (observed frame) map
    with a limit of 26 R-\marc~(bottom) for RXCJ\,0014.3$-$3022\,NW.
    R-band isophotes highlight different objects.
    Bright galaxies as red as the BCGs are at $z \sim 0.3$
    (Braglia et al. 2007).}
  \label{FigBR1a}
\end{figure*}

\begin{figure*}
  \centering
  \includegraphics[width=15cm]{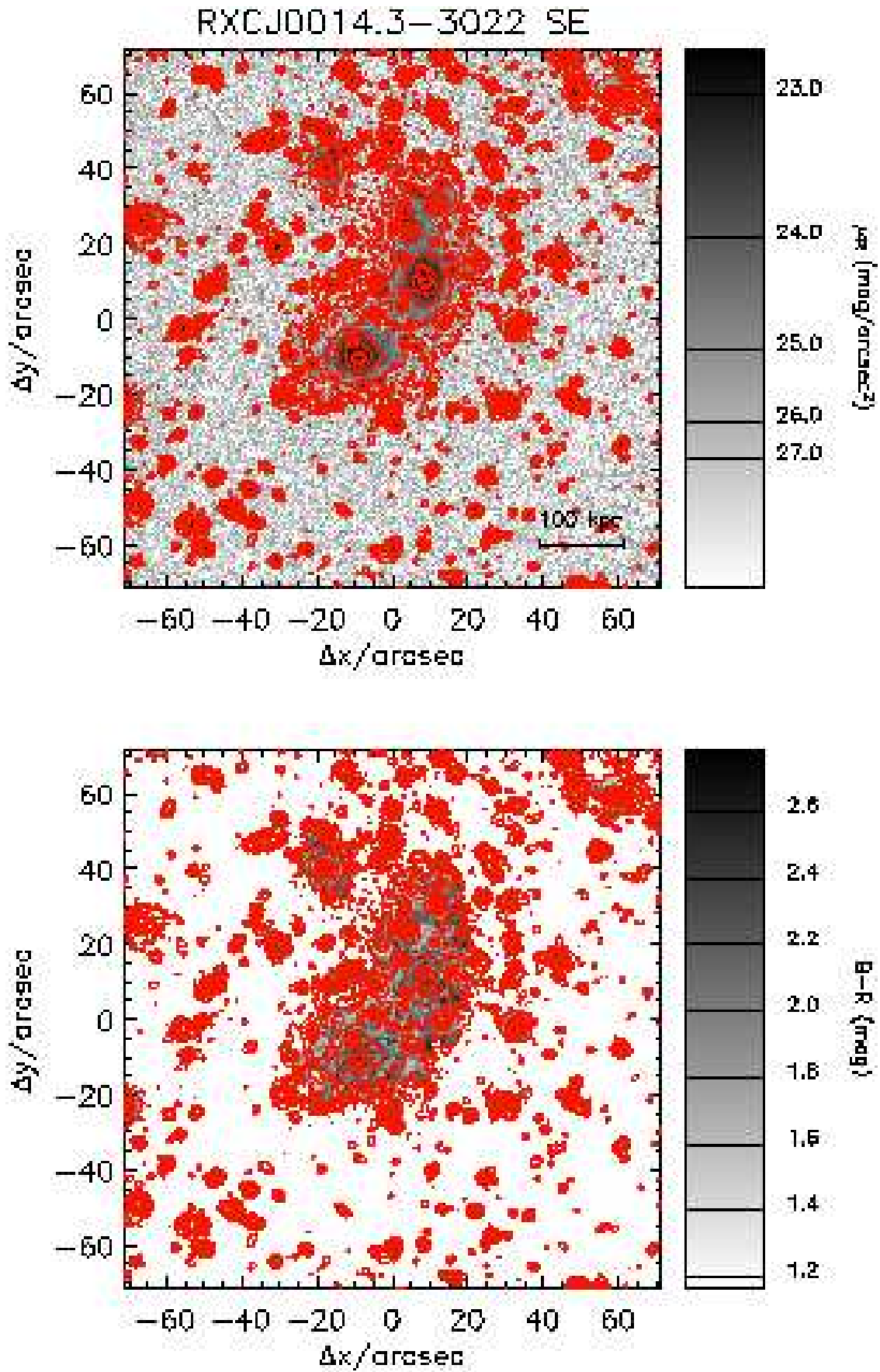}
  \caption{The same as in Fig.~\ref{FigBR1a} but for the SE pair of BCGs
    in RXCJ\,0014.3$-$3022.
    Bright galaxies as red as the BCGs are at $z \sim 0.3$
    (Braglia et al. 2007).}
  \label{FigBR1b}
\end{figure*}

\begin{figure*}
  \centering
  \includegraphics[width=15cm]{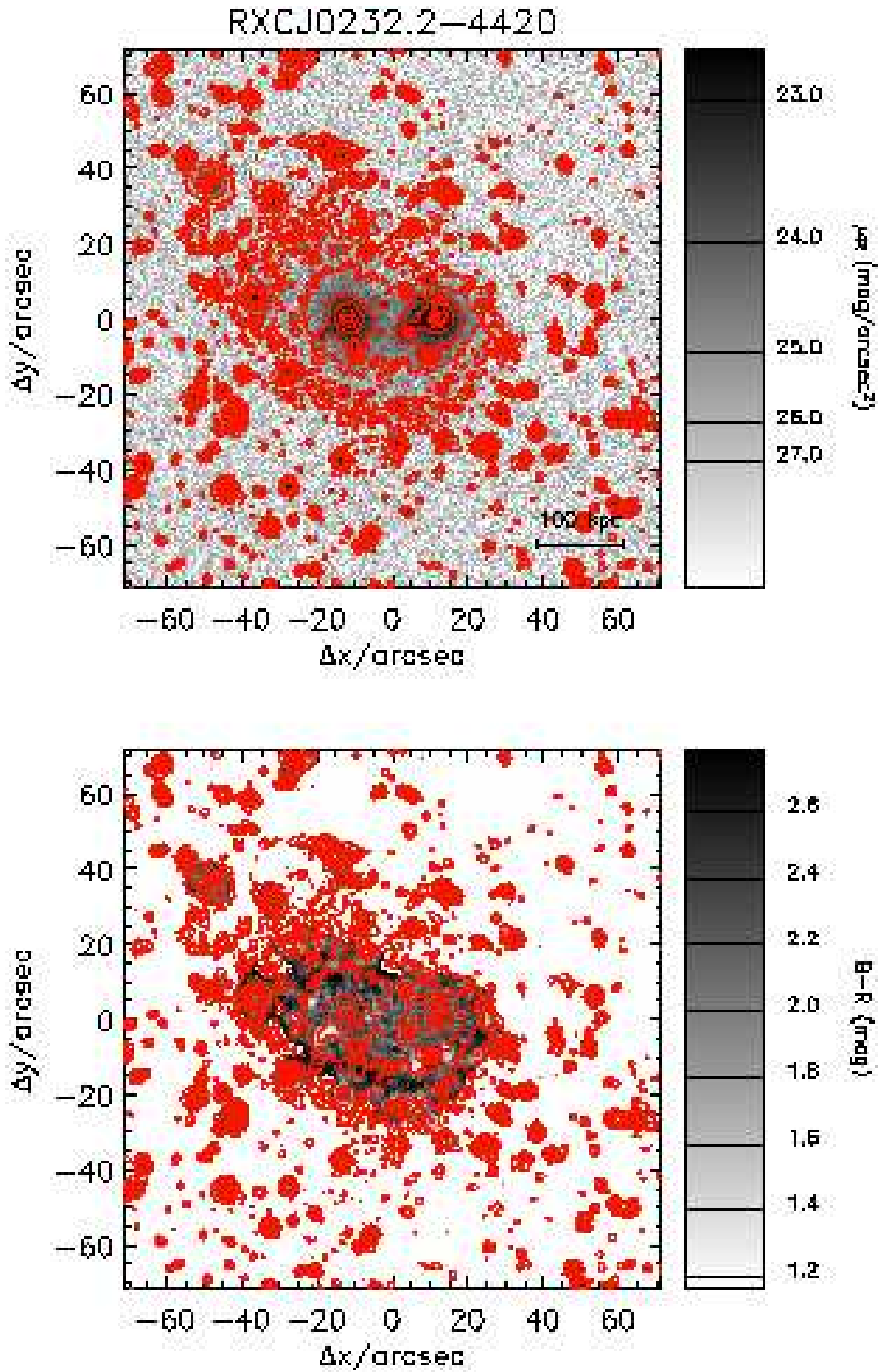}
  \caption{The same as in Fig.~\ref{FigBR1a} but for the pair of BCGs
    in RXCJ\,0232.2$-$4420.
    Bright galaxies as red as the BCGs are likely at $z \sim 0.3$,
    as confirmed for RXCJ\,0014.3$-$3022 (Braglia et al. 2007).}
  \label{FigBR2}
\end{figure*}

\begin{figure*}
  \centering
  \includegraphics[width=15cm]{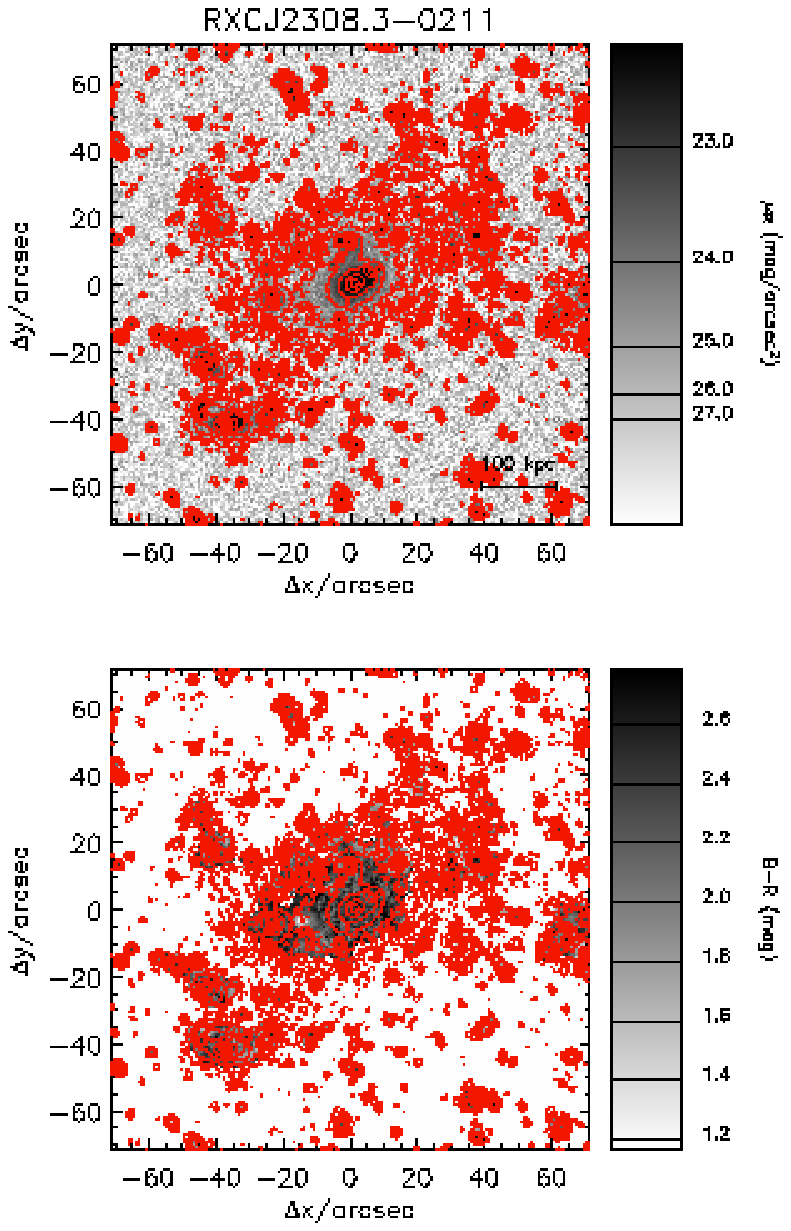}
  \caption{The same as in Fig.~\ref{FigBR1a} but for the single BCG
    in RXCJ\,2308.3$-$0211.
    Bright galaxies as red as the BCG are likely at $z \sim 0.3$,
    as confirmed for RXCJ\,0014.3$-$3022 (Braglia et al. 2007).}
  \label{FigBR3}
\end{figure*}

The upper panels of Figs.~\ref{FigBR1a}--\ref{FigBR3} show zoom-in's
of the R-band images of the clusters, centred on their (pairs of) BCGs.
Isophotal contours are drawn down to 26~\marc;
they clearly show diffuse luminous envelopes around all seven BCGs.
Diffuse stellar emission is detected down to 26 R-\marc~not only
around the BCGs but also around other bright member galaxies,
as in RXCJ\,0014.3$-$3022\footnote{Membership can be assigned
from the spectroscopic and/or photometric redshifts determined
in Braglia et al. (2007).}.
However, we keep our focus on the core regions of galaxy clusters
and refer the reader to Krick \& Bernstein (2007) for a discussion
of the ICL forming earlier in infalling groups
and later combining in the cluster centre.

\begin{landscape}
\begin{figure}
  \centering
  \includegraphics[angle=270,width=20cm]{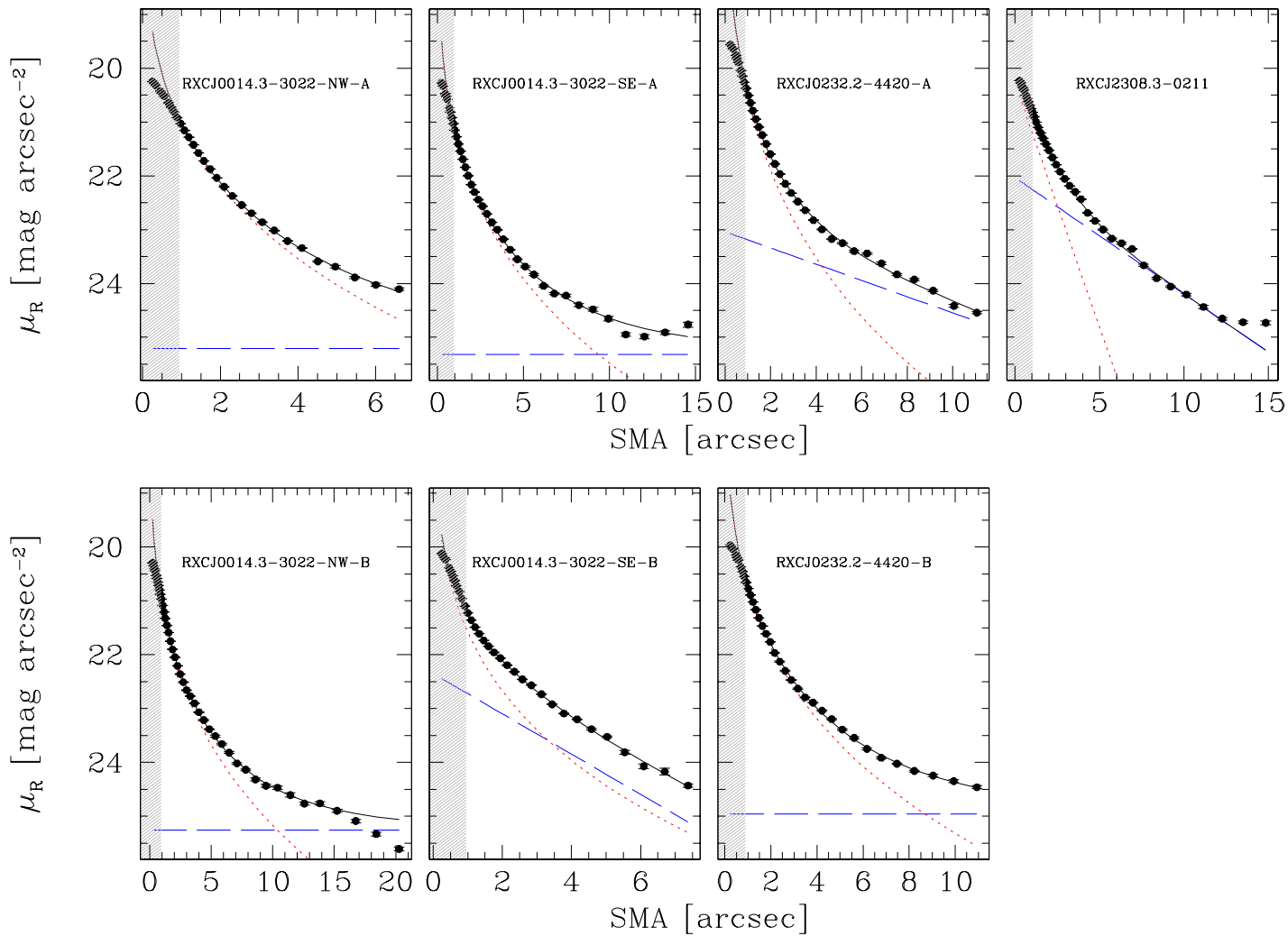}
  \caption{R-band radial surface brightness profiles of the BCGs.
    In each panel, inner ``de Vaucouleurs'' (except for the BCG
    in RXCJ\,2308.3$-$0211) and outer exponential components
    are marked with dotted and long-dashed lines, respectively,
    whereas a solid line represents the total best-fit model.
    A grey shaded area marks the innermost region not used
    for the surface brightness profile fitting because affected by seeing.
    ``SMA'' stands for semi-major axis of the individual elliptical annuli
    (see text).}
  \label{SBprofs}
\end{figure}
\end{landscape}

The surface brightness distribution of the seven BCGs is further analysed
by means of elliptical isophote fitting
(through the IRAF\footnote{IRAF is distributed
by the Na\-tio\-nal Opti\-cal Astro\-nomy Observa\-tories, which are operated
by the Association of Universities for Research in Astronomy, Inc.,
under cooperative agreement with the National Science Foundation.}
task \texttt{ellipse}) and azimuthally averaged surface brightness profiles.
This is reported in Fig.~\ref{SBprofs},
where data points are logarithmically spaced.
Other galaxies are conservatively masked out
before elliptical isophotes are fitted.
We take a similar approach as Seigar et al. (2007)
and decompose each profile in two components:
an inner ``de Vaucouleurs'' $r^{1/4}$-law (de Vaucouleurs 1948)
and an external exponential component.
The only exception is the BCG of RXCJ\,2308.3$-$0211,
for which a double exponential represents a better fit.
Fits are performed assigning the same weight to each data point
in order to properly weigh the outer parts of the profiles.
These would be highly under-weighted
in a scheme based on the signal-to-noise ratio (S/N).
Furthermore, points at semi-major axis smaller than the seeing FWHM
are excluded.
We stress, however, that the actual analytic model or weighting scheme adopted
is not crucial for the qualitative arguments discussed below.

For five out of seven BCGs (i.e., RXCJ\,0014.3$-$3022\,NW\,A and B,
RXCJ\,0014.3$-$3022\,SE\,A, and RXCJ\,0232.2$-$4420\,A and B),
the best-fit decomposition is given by an inner ``de Vaucouleurs''
plus a very flat outer exponential.
This is suggestive of a ``classical'' elliptical core
living in a very diffuse halo
with $\mu_{\mathrm R} \gtrsim 25$~\marc~(cf. e.g. Schombert 1987).
A single ``S\'ersic'' profile (S\'ersic 1968) provides a much worse fit
in all these five cases.
For the other two BCGs, the inner profile is more complex
than a single ``de Vaucouleurs'' law and the contamination by nearby galaxies
does not enable us to extend the profile as deep as to reach
the alleged surface brightness limit where the diffuse component takes over.
We note that these results are in reasonable agreement
with those of Seigar et al. (2007), based on a sample of five nearby cDs.

As the faintest R-band isophotes drawn in Figs.~\ref{FigBR1a}--\ref{FigBR3}
indicate, the detected diffuse stellar emission extends
up to $\sim 100~\mathrm{kpc}$ from the (closest) BCG.
This extent is well beyond the (projected) R-band effective radius
of individual BCGs, which measures $\sim 7.5$--$27.4~{\mathrm{kpc}}$
(see $r_{\mathrm{e,R}}$ in Table~\ref{BCGProp}).
This gives us confidence that a non-negligible fraction
of the light detected at the largest distances from the individual BCGs
is emitted by stars that are free-floating in the cluster
rather than gravitationally bound to a BCG.
As a comparison, Zibetti et al. (2005) find that
on average 30--40\% of the emission detected at such large distances
is due to the intracluster component.

The detection of diffuse light in our images is robust
in spite of their relatively short exposures (cf. Krick \& Bernstein 2007).
In fact, the 26~R-\marc~isophotes drawn in Figs.~\ref{FigBR1a}--\ref{FigBR3}
are significant at the 3--5.7$\sigma$ level by assuming that
background fluctuations represent the main source of uncertainty.
As a further proof of the reliability of our detections,
we confirm the presence of diffuse stellar emission in RXCJ\,0014.3$-$3022
at the same locations where it was discovered
by Krick \& Bernstein (2007).
%
%__________________________________________________________________

\subsection{Diffuse stellar emission and intracluster medium}\label{distrib}

If a relevant fraction of the diffuse stellar emission around BCGs
is not bound to any BCG, the projected distribution of this component
can be expected to deviate from the isophotal shape of the BCG(s)
and resemble the projected distribution of the cluster's DM halo.
This is mapped by the X-ray emitting gas in relatively relaxed clusters
(e.g. Allen et al. 2002).

The distribution of the diffuse light
around the pair of BCGs in RXCJ\,0232.2$-$4420
resembles that of the envelopping X-ray emitting gas.
The centres of these two BCGs are aligned in the EW direction
and separated by $\sim 100~{\mathrm{kpc}}$;
the major axes of RXCJ\,0232.2$-$4420\,A and B have position angles
(measured counter-clockwise from north to east)
of about 45\degr~and 100\degr, respectively (see Fig.~\ref{FigBR2}).
The cocoon of diffuse stellar emission within the 26 R-\marc~isophote
exhibits a SW extention which reaches out to $\sim 80~{\mathrm{kpc}}$
towards the west at a galactocentric distance of $\sim 90~{\mathrm{kpc}}$
along the direction of the major axis of RXCJ\,0232.2$-$4420\,A.
This behaviour is analogous to that of the X-ray emission of the cluster,
whose projected distribution is centred on RXCJ\,0232.2$-$4420\,A
but deviates from centrally symmetry owing to a broad SW extention
(see Fig.~\ref{FigXrayOpt2}).

RXCJ\,2308.3$-$0211 exhibits some correspondence between the distributions
of diffuse light and X-ray emission (see Fig.\ref{FigXrayOpt3}).
In this fairly regular cluster, there is a single BCG
and the X-ray emission peak coincides with its position.
Thus, an alignment between the major axes of the BCG
and the potential well of its parent cluster could be expected
(Brown 1997; Dubinski 1998 and references therein).
The ``peculiar'' R-band radial surface brightness profile
of the BCG in RXCJ\,2308.3$-$0211 (see Fig.~\ref{SBprofs})
may witness the presence of diffuse light not bound to the BCG.

The ICM does not necessarily trace DM in a merging cluster
like RXCJ\,0014.3$-$3022 (cf. Clowe et al. 2006).
There diffuse light is detected around two pairs of BCGs,
which probably experience severe gravitational stresses
also owing to the ongoing large-scale merging.
In particular, RXCJ\,0014.3$-$3022\,NW suggests strong tidal forces in action:
a stellar ``plume'' stretches out of RXCJ\,0014.3$-$3022\,NW\,B
along the bridge of diffuse emission connecting the two BCG
main bodies - as defined by their effective radii - (see Fig.~\ref{FigBR1a}).
In absence of spectroscopic evidence of interaction,
the stellar plume stretching out of RXCJ\,0014.3$-$3022\,NW\,B
and the twisted outer isophotes of RXCJ\,0014.3$-$3022\,NW\,A
support the hypothesis that the stellar bridge between these BCGs
is not a projection effect.
%
%__________________________________________________________________

\subsection{Colour of the diffuse stellar emission around BGCs}\label{col}

Disentangling the contributions of the ICL and BCG
to the observed diffuse stellar emission around a BCG
can be hampered by galaxy crowding in a cluster core
(see Krick \& Bernstein 2007 for RXCJ\,0014.3$-$3022).
However, a colour analysis may constrain the properties of intracluster stars
if the ICL makes on average 30--40\% of the total optical emission
(including galaxies) at $\sim 100~\mathrm{kpc}$ (Zibetti et al. 2005).
Therefore, we produce colour maps of our fields and investigate
whether the expected presence of intracluster stars
of as-yet unkown origin and amount affects the colours of the diffuse light
around our sample BCGs.
First we adaptively median smooth the images, in both R and B bands.
The intensity of each pixel is replaced with the median intensity
in a circle whose radius is grown until a minimum S/N is reached.
Given the properties of our imaging, we require a minimum S/N of 20 in R,
and of 5 in B.
The same smoothing is applied in both bands;
it results in $\mathrm{B-R}$ colour uncertainties smaller than 0.25 mag
on all regions with R-band surface brightness
$\mu_{\mathrm{R}} \la 26$~\marc.
This scheme preserves the full original spatial information and resolution
in the highest S/N regions (i.e., within the ``classical'' optical extent
of a galaxy).
It also allows optimal S/N measurements of surface brightness and colour
to be obtained in low surface brightness regions, such as galaxy outskirts
and intergalactic regions.
$\mathrm{B-R}$ colour maps
are obtained from the adaptively median-smoothed images;
they are shown in the bottom panels of Figs.~\ref{FigBR1a}--\ref{FigBR3}.

\begin{figure*}
  \centering
  \includegraphics[width=15cm]{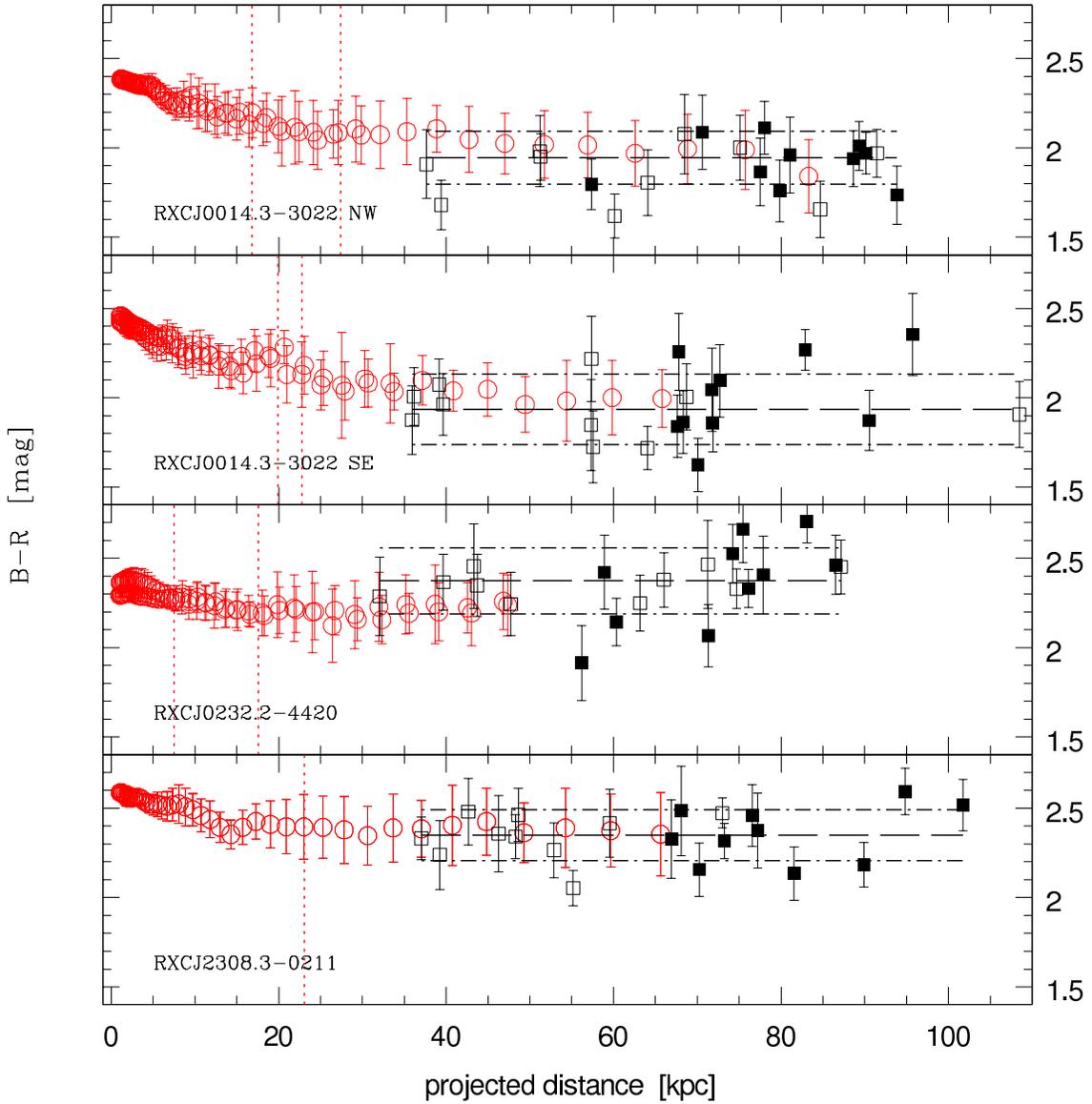}
  \caption{$\mathrm{B-R}$ colours (observed frame) as measured
    at different projected radial distance from the centre
    of the (closest) BGC for the three pairs of BCGs RXCJ\,0014.3$-$3022\,NW,
    RXCJ\,0014.3$-$3022\,SE, RXCJ\,0232.2$-$4420 plus the single BCG
    in RXCJ\,2308.3$-$0211 listed in Table~\ref{BCGProp}.
    In each panel, empty circles reproduce the radial $\mathrm{B-R}$
    colour profile of each BCG measured in elliptical annuli;
    the RMS of the colour in each annulus is shown as an error bar.
    A vertical dotted line indicates the R-band effective radius of the BCG.
    Empty squares represent median values of the diffuse light
    estimated in 10 different 81 square pixel-regions
    (i.e., areas of $\sim 90~\mathrm{kpc}^2$ at $z=0.3$)
    that are free from high-surface brightness emission
    associated with non-BCG galaxies and located all around
    each (pair of) BCG(s) but close to the 26 R-\marc~isophote.
    Filled squares represent median values of the diffuse light
    estimated in 10 different 81 square pixel-regions
    that are free from high-surface brightness emission
    associated with non-BCG galaxies and located at a projected distance
    of more than $50~\mathrm{kpc}$ from any BCG.
    Error bars represent values of the standard deviation
    of the $\mathrm{B-R}$ colour distribution
    as probed in individual regions of the colour maps.
    Horizontal long-dashed and dot-dashed lines mark, respectively,
    the median and the median $\pm$ 1 RMS values of the colour
    of the diffuse stellar emission (see text).}
  \label{FigCol}
\end{figure*}

Given the nature of the diffuse light and the depth of our photometry,
a statistics of the $\mathrm{B-R}$ colour of the diffuse stellar emission
around BCGs is computed in 20 regions, each of 4.6 square arcsec
(i.e., $9 \times 9$ square pixels), on individual colour maps.
Ten of these regions are located all around each (pair of) BCG(s)
but close to the 26 R-\marc~isophote;
10 are located at a projected distance larger than $50~\mathrm{kpc}$
from the centre of the (closest) BCG.
The 20 areas avoid high-surface brightness regions
that are clearly associated with other galaxies than the BCGs.
They can partially overlap, especially those at the largest distances
from the (closest) BCG.
In this approach, we consider as the characteristic $\mathrm{B-R}$ colour
of the diffuse light around each (pair of) BCG(s) the median of the 20 values
of the median $\mathrm{B-R}$ colour estimated in the different
peripheral regions of each (pair of) BCG(s).
This choice is motivated by the absence of colour gradients
at large distances from the (closest) BCG (see Fig.~\ref{FigCol}).
It also minimizes the impact of contaminating emission
at low surface brightness from bright galaxies other than the BCGs,
galaxies with very low-surface brightnesses or faint, unresolved sources.
For BCGs, total colours $(\mathrm{B-R})_{\mathrm{tot}}$ are computed
from the B and R magnitudes listed in Table~\ref{BCGProp},
whose accuracy mostly depends on source deblending
(see Figs.~\ref{FigBR1a}--\ref{FigBR3}).
In addition, we determine radial $\mathrm{B-R}$ colour profiles
for the individual BCGs.
Elliptical annuli are defined from the R-band elliptical isophotes
(Sect.~\ref{detect}), whereas average $\mathrm{B-R}$ colours
inside annular regions are derived from the colour maps.
Figure~\ref{FigCol} reproduces the $\mathrm{B-R}$ colour
of the diffuse light - as measured in individual regions
at distances larger than $30~\mathrm{kpc}$ from the centre
of the (closest) BCG - (squares) and the $\mathrm{B-R}$ colour profile
of the associated BCG(s) (circles) for each (pair of) BCG(s).
No $k$- nor evolutionary correction is applied to our photometry:
they critically depend on the assumption
made on the spectral energy distributions of the stellar populations,
whose nature is a priori unknown.

In all the seven BCGs the $\mathrm{B-R}$ colour
is equal to $\sim 2.3$--2.6 mag in the central region and becomes bluer
at the effective radius, on average by $\sim 0.2~\mathrm{mag}$
(see Table~\ref{BCGICLCol}).
This is consistent with observations of the local Universe:
colours are remarkably constant within the inner regions of BCGs
(Postman \& Lauer 1995), and strong colour gradients are on average absent
in cD galaxies (e.g., Schombert 1988; Mackie 1992; see however Brown 1997).
In the rather relaxed clusters RXCJ\,0232.2$-$4420 and RXCJ\,2308.3$-$0211
the diffuse light around BCGs exhibits the same colour,
$\mathrm{B-R} \sim 2.4$, which is consistent at the 1--2$\sigma$ level
with the colours measured within the effective radii of the individual BCGs.
Conversely, in the merging cluster RXCJ\,0014.3$-$3022
the diffuse stellar emission around each (pair of) BCG(s)
exhibits $\mathrm{B-R} \sim 1.9$, which is $0.35 \pm 0.12~\mathrm{mag}$ bluer
than the average total colour $\mathrm{B-R} \sim 2.3$ of the four BCGs,
and, a fortiori, of their central regions\footnote{Krick \& Bernstein (2007)
estimate the colour of the diffuse light in RXCJ\,0014.3$-$3022
to be $\mathrm{V-r} \simeq 1.0 \pm 0.8$, which is significantly redder
(0.6 mag) than the cluster red sequence that they determine. Clearly
our result is at odds with theirs.}.
The colour profile of one BGC in each pair
(i.e., RXCJ\,0014.3$-$3022\,NW\,B and RXCJ\,0014.3$-$3022\,SE\,A)
extends out to at least $65~\mathrm{kpc}$, where its value is consistent
with the colour of the diffuse light further out.
%
%__________________________________________________________________

\section{Discussion}\label{disc}

The existence of a diffuse population of intergalactic stars
in galaxy clusters originating the ICL is well-established observationally,
but several interpretations of the ICL are possible (see Sect.~\ref{intro}).
In particular, there are observations and simulations which indicate
that the formation of the diffuse stellar component around a BCG
is mostly associated with the build-up of the BCG itself
(Zibetti et al. 2005; Murante et al. 2007).
The present understanding of the hierarchical formation of BCGs
(De Lucia \& Blaizot 2007) foresees that their stars are formed very early
(50\% at $z \sim 5$, 80\% at $z \sim 3$) and in many small galaxies.
Model BCGs assemble surprisingly late: half their final mass
is typically locked up in a single galaxy after $z \sim 0.5$.
These late mergers do not change the apparent age of BCGs,
since they involve the accumulation of a large number
of old stellar populations from galaxies with little gas content
and red colours.
This yields the observed homogeneity of BCG properties in the local Universe
(e.g., Schombert 1988; Mackie 1992; Postman \& Lauer 1995;
von der Linden et al. 2007; see however Brown 1997).

Different observational results from samples of clusters at different redshifts
do not lead to a unanimous consensus on this theoretical scenario.
For instance, the K-band Hubble diagram for 25 BCGs at $0 < z < 1$
suggests that the stellar mass in a typical BCG has grown in good agreement
with the predictions of semi-analytic models of galaxy formation and evolution
set in the context of a hierarchical scenario for structure formation
(Aragon-Salamanca et al. 1998).
On the other hand, the lack of evidence for evolution
of the central-galaxy luminosity--host-halo mass relation
for a sample of known BCGs at $0.1 < z < 0.8$ points to a BCG growth
that is still limited by the timescale for dynamical friction
at these earlier times, not proceeding according to the predictions
of current semi-analytic models (Brough et al. 2008).

Observations of BCGs at intermediate redshifts
are crucial for establishing the late assembly of BCGs
and understanding the origin of the ICL.
Thus, we selected three clusters from the REFLEX-DXL sample (Zhang et al. 2006)
with available wide-field optical imaging (Pierini et al. in preparation),
i.e.: RXCJ\,0014.3$-$3022, RXCJ\,0232.2$-$4420, and RXCJ\,2308.3$-$0211.
This sample is limited but spans the range of dynamical states
of the most X-ray luminous clusters at $z \sim 0.3$.
We cannot tackle important issues such as the evolution of the ICL fraction
and colours (see Zibetti et al. 2005; Krick \& Bernstein 2007)
for lack of statistics and redshift coverage.
However, we demonstrate that medium--deep (down to a completeness limit
of $\mathrm{R} \sim 23.5$), multi-band (B and R bands here) photometry
at a 2m-class telescope can foster the knowledge of the ICL
in individual clusters at $z \sim 0.3$.
In fact, we robustly detect diffuse stellar emission
down to 26 R-\marc~(observed frame) with 1 hr-exposures.
The depth of our photometry enables us to investigate
distribution and colour of the diffuse light around BCGs
as a function of the dynamical state of a galaxy cluster.

Diffuse stellar emission is detected around all the seven BCGs
in the three clusters, and around other bright galaxies in RXCJ\,0014.3$-$3022
(cf. Krick \& Bernstein 2007).
This diffuse light exceeds the emission from stars
with a projected distribution following a $r^{1/4}$-law,
and extends up to $\sim 100~\mathrm{kpc}$ from the centre of the (closest) BCG.
At this distance, we expect that
on average 30--40\% of the diffuse stellar emission is due to the ICL,
as from the statistical results obtained from 683 clusters of galaxies
at $0.2 \la z \la 0.3$ (Zibetti et al. 2005).
Consistently, the distribution of the diffuse light around BCGs
follows the larger-scale distribution of the ICM
in a system which is either relaxed (i.e., RXCJ\,2308.3-0211)
or close to relaxation (i.e., RXCJ\,0232.2-4420).
There the ICM is expected to trace the DM component,
which dominates the mass budget of a cluster.
In absence of dynamical information, this is indirect evidence that
part of the diffuse stellar emission around three out of seven BCGs
is associated with the potential well of their parent clusters.

\begin{landscape}
\begin{figure}
  \centering
  \includegraphics[width=24cm]{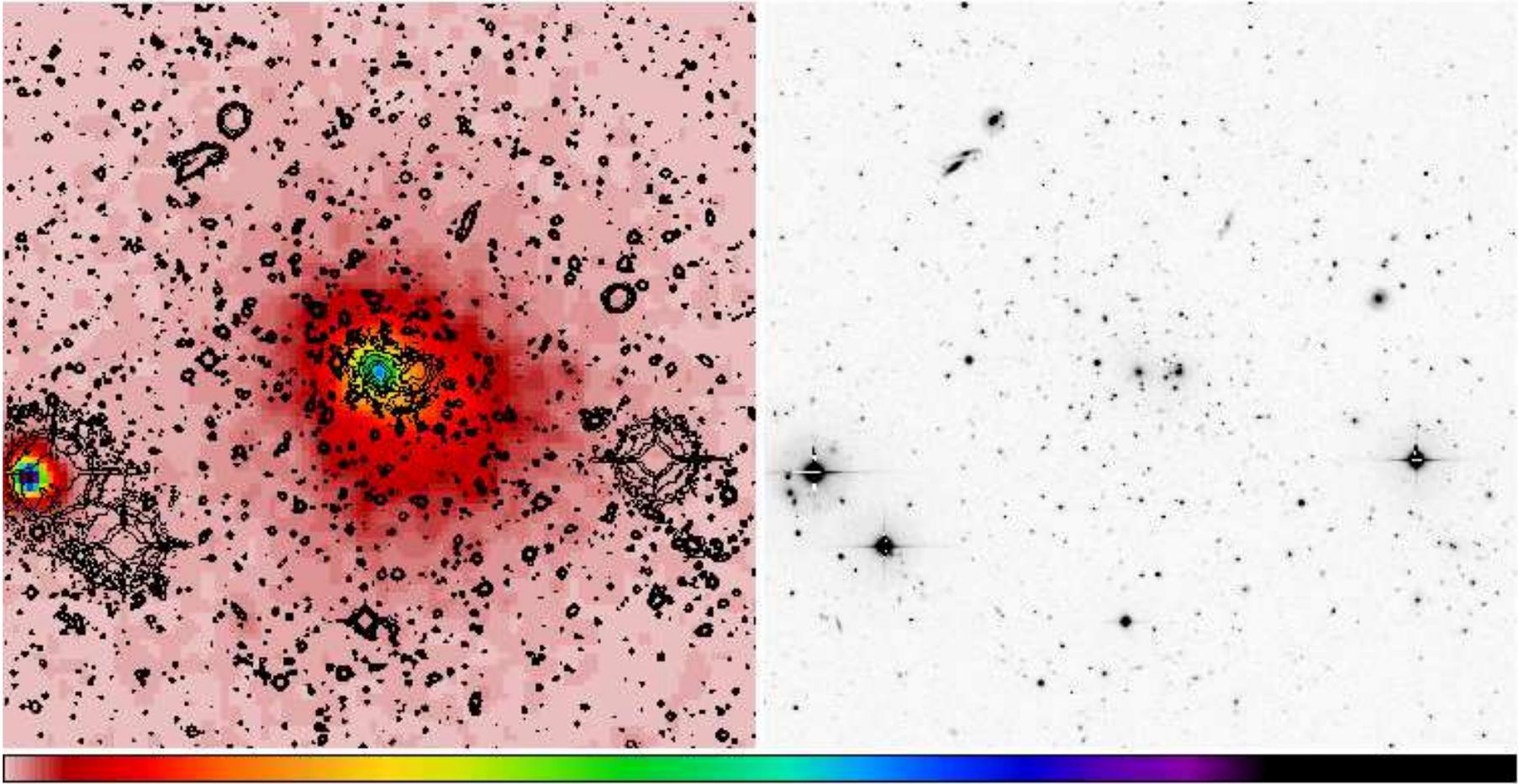}
  \caption{RXCJ\,0232.2$-$4420 as imaged by EPIC and WFI.
    Size and symbols are the same as in Fig.~\ref{FigXrayOpt1}.
    The limiting R-band surface brightness is equal to 25.81~\marc.
    Note the presence of a foreground group of late-type galaxies
    in the upper half of the R-band image.
    No detected X-ray emission appears to be associated
    with this galaxy group.}
  \label{FigXrayOpt2}
\end{figure}
\end{landscape}

Comparison of Figs.~\ref{FigXrayOpt1}--\ref{FigXrayOpt3} suggests that
BCGs and diffuse light around them have distinct dynamical evolutions.
This is in agreement with the conclusion of Krick \& Bernstein (2007).
The diffuse light around the two pairs of BCGs in RXCJ\,0014.3$-$3022
\emph{can be at least in part the result of severe gravitational stresses
potentially experienced by these BCGs\/} (see Sect.~\ref{distrib}).
Each pair of BCGs, offset by $\sim 210~\mathrm{kpc}$
from the closest X-ray subcomponent of this merging cluster,
probably signposts the minimum of the potential well
of each dark matter subcomponent (cf. Clowe et al. 2006).
The presence of a pair of BCGs at the centre of RXCJ\,0232.2$-$4420
together with the X-ray morphology (Fig.~\ref{FigXrayOpt2}) suggest that
this system is evolving towards relaxation, like Coma.
The diffuse stellar emission detected at the largest distances
from this pair of BCGs follows the distribution of the dark matter
(i.e., the X-ray emitting, hot gas), \emph{whether the intergalactic stars
were stripped from the BCGs or had a different origin\/}
(see Krick \& Bernstein 2007).
One can imagine that, later on,
the two BCGs will merge into a new dominant galaxy
(e.g., Naab et al. 2006; De Lucia \& Blaizot 2007),
which will sit at the centre of the cluster potential well,
and possibly respond to its triaxial configuration.
Eventually the cluster will look like RXCJ\,2308.3$-$0211,
where the distribution of the diffuse light around the single BCG
in the cluster core is aligned with the major axes of the BCG
and of the larger-scale distribution of the ICM (i.e., the dark matter).

All seven BCGs exhibit $\mathrm{B-R}$ colours
which differ by up to 0.3 mag in their central regions.
A similar narrow spread exists among their colour gradients,
when $\mathrm{B-R}$ colours are measured in the central region
and at the effective radius of each BCG (see Table~\ref{BCGICLCol}).
These results are consistent with others in the literature
(e.g., Schombert 1988; Mackie 1992; Postman \& Lauer 1995).
They confirm that star formation has occurred over a short time-scale
in the BCGs under study (cf. von der Linden et al. 2007
and Bildfell et al. 2008).
Furthermore, the $\mathrm{B-R}$ colour of the diffuse light
around the BCGs in RXCJ\,0232.2$-$4420 and RXCJ\,2308.3$-$0211 is about as red
as the light within the effective radii of the BCGs (see Fig.~\ref{FigCol}),
whether there is a pair of BCGs or a single BCG
in the cluster core\footnote{Isolation and uniqueness of the BCG
may have led to the redder core of the dominant central galaxy
in RXCJ\,2308.3$-$0211, possibly owing to a lack of recent gas accretions
(cf. Bildfell et al. 2008).}.
This is consistent with the result of Zibetti et al. (2005).
Conversely, the $\mathrm{B-R}$ colour of the diffuse light
detected around the two pairs of BCGs in RXCJ\,0014.3$-$3022
is bluer than the average colour of the central regions
of the two pairs of BCGs by $0.44 \pm 0.15$--$0.51 \pm 0.20$ mag.

\begin{figure*}
  \centering
  \includegraphics[width=12cm]{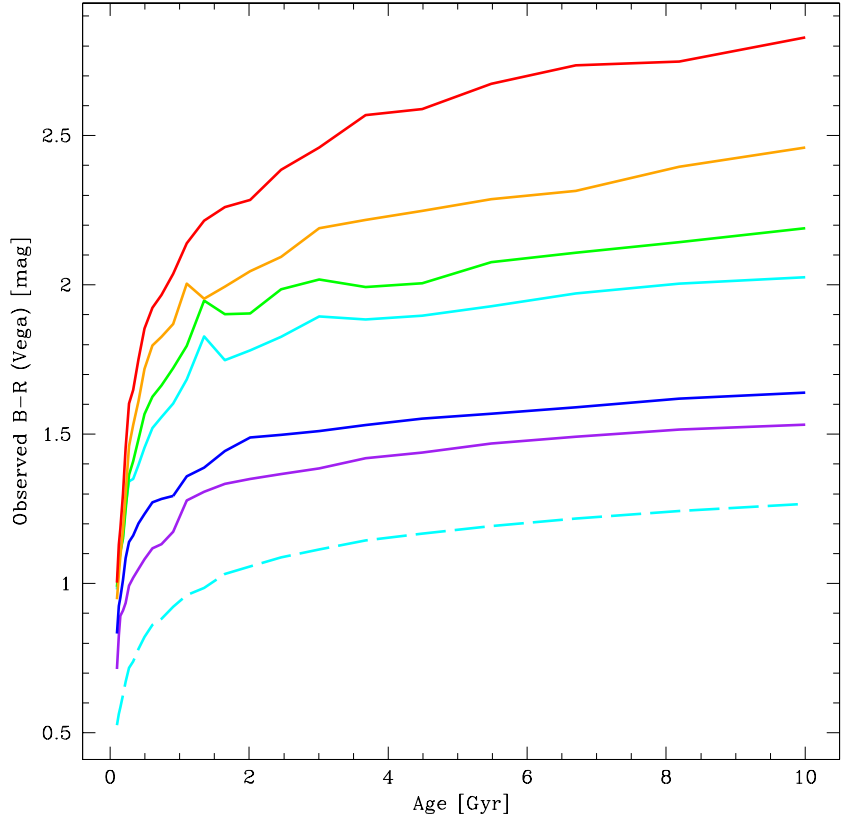}
  \caption{$\mathrm{B-R}$ colours as a function of the epoch
    since star formation started (i.e., the ``age'') and metallicity
    for models with different star-formation histories
    (Bruzual \& Charlot 2003) redshifted to $z=0.3$.
    From top to the bottom, solid curves
    reproduce simple stellar population models (i.e., models
    of a single burst of star formation) with fixed metallicity $Z$
    equal to 2.5, 1, 0.4, 0.2, 0.02, and $0.005 \times \mathrm{Z}_{\sun}$,
    respectively.
    The long-dashed curve reproduces models
    with constant star formation rate and $Z = 0.2 \times \mathrm{Z}_{\sun}$.
    All models share the same stellar initial mass function (Chabrier 2003).}
  \label{FigBC03Mod}
\end{figure*}

In analogy with early-type galaxies at different redshifts
(Saglia et al. 2000; Tamura et al. 2000),
mild colour gradients can imply mild metallicity gradients in our BCGs.
We make use of the stellar population evolutionary synthesis models
in Bruzual \& Charlot (2003) to interprete the observed $\mathrm{B-R}$ colours
of BCGs and diffuse light.
Comparison between Fig.~\ref{FigCol} and Fig.~\ref{FigBC03Mod} shows
that the colours measured in the central regions of all the seven BCGs
can be reproduced by simple stellar population (SSP)
models with at least solar metallicity ($Z_{\sun}$)
and age older than $2~\mathrm{Gyr}$\footnote{The interpretation
of broad-band colours suffers from the age--metallicity degeneracy
(Worthey 1994).}.
The colours measured at the effective radius of each BCG
can be reproduced with SSP models spanning the same range in age
and the range $\sim 0.4$--$1 \times Z_{\sun}$ in metallicity.
Furthermore, the diffuse stellar component around the BCGs
in RXCJ\,0232.2$-$4420 and RXCJ\,2308.3$-$0211
has metallicity larger than solar and is probably as old
as the central regions of the associated BCG(s).
Conversely, the colour of the diffuse light around the two pairs of BCGs
in the merging cluster RXCJ\,0014.3$-$3022 can be reproduced with SSP models
with age between 0.5 and $4~\mathrm{Gyr}$ and metallicity higher than
$\sim 0.2 \times Z_{\sun}$.
Hence, \emph{there is a difference in the stellar populations
responsible for the diffuse light around the BCGs under study\/}.

The stellar population originating the bluer diffuse light
around the two pairs of BCGs in RXCJ\,0014.3$-$3022 can be of internal origin.
Under this hypothesis, it is either as old as the stellar populations
within the effective radii of the individual BCGs but of lower metallicity,
or associated with recent star-formation activity\footnote{This holds
whether the likely interactions of the two pairs of BCGs
in RXCJ\,0014.3$-$3022 were triggered by the large-scale merging
or started in the two subcomponents of the system before this event.}.
In the first case, the star formation histories of the four BCGs
in RXCJ\,0014.3$-$3022 had to be rather similar,
although the two pairs of BCGs inhabited distinguished environments
well before the epoch of observation (cf. Sect.~\ref{notes}).
This can not be excluded on the basis of the available data but is suspicious.
On the other hand, in a sample of 48 X-ray luminous galaxy clusters
at intermediate redshifts, colour profiles that turn bluer
towards their centres are held as evidence for recent star formation in BCGs
(e.g., Bildfell et al. 2008).
Likely the stars responsible for the bluer colour
of the diffuse light around the two pairs of BCGs in RXCJ\,0014.3$-$3022
were not originated in the BCGs.

\begin{figure*}
  \centering
  \includegraphics[width=18cm]{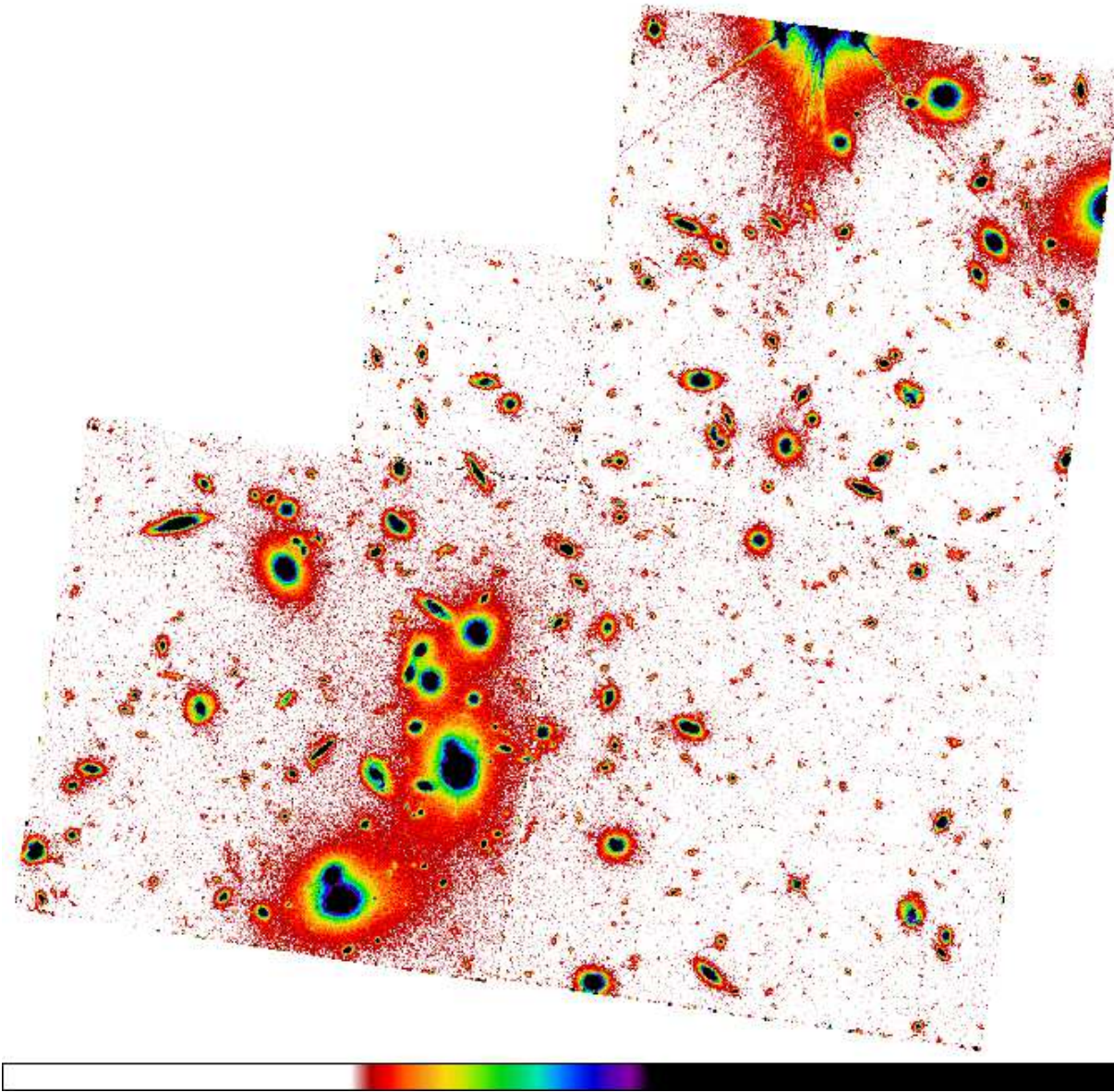}
  \caption{Archival image of the RXCJ\,0014.3$-$3022 cluster core
    at a spatial resolution of $0.3^{\prime \prime}$
    (north is up and east to the left).
    It was obtained from a 36 min-exposure in the F702W filter - analogous
    to the R-band filter - with the Wide Field Planetary Camera
    (WFPC2) of the Hubble Space Telescope (HST) (see Couch et al. 1998).
    This reproduction highlights galaxy features
    at low surface brightness with a linear scale where intensity increases
    as the colour turns from white to black.}
  \label{wfpc2}
\end{figure*}

This external stellar population may simply be part of unidentified,
low surface brightness galaxies that are rather homogeneously distributed
around the two pairs of BCGs in RXCJ\,0014.3$-$3022
out to galactocentric distances of $\sim 100~\mathrm{kpc}$ (unlikely).
If the parent galaxies are faint and unresolved in the WFI images,
the FWHM of their R-band surface brightness distribution
must be $\leq 1^{\prime \prime}$, which corresponds to $\leq 4.5~\mathrm{kpc}$
at $z \sim 0.31$.
However, imaging of RXCJ\,0014.3$-$3022\,SE at comparable depth
with respect to the WFI image but at three times-higher angular resolution
does not reveal the presence of a significant population of galaxies
which are not identified in the latter image
(cf. Fig.~\ref{wfpc2} and Fig.~\ref{FigBR1b}).
Thus, the bluer colour of the diffuse light in RXCJ\,0014.3$-$3022
may indeed be due to a diffuse stellar component:
the intergalactic stars are either produced in situ (unlikely)
or come from shredding of galaxies that are bluer than the BCGs
(favoured scenario).
The longer the duration of the star-formation activity in the galaxies
of origin, the lower the mass fraction of these external blue stars
(cf. the colours for single burst and continuous star formation models
with $0.2 \times Z_{\sun}$ in Fig.~\ref{FigBC03Mod}).
On the other hand, as the mass fraction of the external blue
stellar population decreases, younger ages and/or lower metallicities
have to be invoked to produce an impact on the observed $\mathrm{B-R}$ colour
of the diffuse light.

A clue to the nature of the shredded galaxies is offered by the study
of ``Butcher--Oemler'' clusters (Butcher \& Oemler 1978) at $z = 0.31$
by Couch et al. (1998)\footnote{The study of star formation rate
and morphology by Couch et al. (1998) is based on ground-based spectroscopy
plus R (F702W) filter observations with the HST-WFPC2 at a spatial resolution
of $0.3^{\prime \prime}$.}, which includes RXCJ\,0014.3$-$3022
(see Fig.~\ref{wfpc2}).
These authors concluded that, on average, one in every five cluster members
(both blue and red ones) shows morphological signatures indicative
of dynamical interactions.
Consistently, Fig.~\ref{wfpc2} shows the presence of several stellar features
reminiscent of tails stemming out of galaxies in the neighbourhood
of RXCJ\,0014.3$-$3022\,SE, as well as a clear asymmetry
in the light distribution of RXCJ\,0014.3$-$3022\,SE\,A.
Dynamical interactions appear to be a common cause
of the most extreme form of star formation activity
seen in the clusters investigated by Couch et al. (1998).
The galaxies involved are mostly of modest luminosity
even in this brightened phase; in their later faded state
they appear destined to become dwarfs.
For the Sdm/Irr Hubble types among the galaxies with ongoing star formation
occurring at the rates typical of normal nearby spirals,
their visibility on the HST-WFPC2 images is due mainly
to their compact, knotty regions of star formation.
These blue, star-forming, dwarf galaxies appear as excellent candidates
to explain the bluer $\mathrm{B-R}$ colour of the ICL in RXCJ\,0014.3$-$3022.
Once they are shredded, the injected stellar populations
would evolve in a passive way - as described by SSPs - and, thus,
become redder at $z=0$ (cf. Krick \& Bernstein 2007).
The gravitational stresses tearing them apart could be excited
also by the ongoing large-scale merging process,
as well as their star formation activity
(cf. the case of cluster MS\,1054$-$03 at $z = 0.83$ in Bai et al. 2007).

Passively evolving, metal poor dwarf galaxies
or star-forming giant galaxies with nearly solar metallicities
can be alternative candidates according to Fig.~\ref{FigBC03Mod}.
Shredding of such galaxies is likely effective
also in the non-merging clusters of our sample
(e.g., Sommer-Larsen et al. 2005; Purcell et al. 2007; cf. Brown 1997),
where the diffuse light is as red as the light within the BCG effective radii;
it may be enhanced in merging clusters, of course.
However, especially if the ICL fraction is lower than average,
it is the shredding of metal poor dwarf galaxies
with temporarily enhanced star-formation activity
the most likely origin of the 0.5 mag-difference in the $\mathrm{B-R}$ colour
between the diffuse stellar emission around BCGs
and the central regions of the BCGs in RXCJ\,0014.3$-$3022.
%
%__________________________________________________________________

\section{Conclusions}\label{conc}

We select three clusters of galaxies
from the ROSAT-ESO Flux-Limited X-ray (REFLEX) cluster survey:
RXCJ\,0014.3$-$3022, RXCJ\,0232.2$-$4420, and RXCJ\,2308.3$-$0211.
They span the morphology range of the most X-ray luminous clusters
at $z \sim 0.3$ observed with XMM-\emph{Newton\/} (cf. Zhang et al. 2006).
Medium--deep, wide-field B- and R-band imaging (Pierini et al. in preparation)
allows us to investigate distribution and colour
of the diffuse stellar emission around the brightest cluster galaxies (BCGs)
as a function of the dynamical state of the parent clusters.
This property is inferred from the X-ray morphology
and optical appearence (e.g., multiplicity of BCGs) of each cluster.

Diffuse stellar emission is robustly detected
around all the seven BCGs in the three X-ray selected clusters,
and its distribution and $\mathrm{B-R}$ colour
down to 26 R-\marc~(observed frame) are investigated.
The diffuse light around BCGs extends
up to $\sim 100~\mathrm{kpc}$ from their centres.
At this distance, we expect that on average 30--40\% of the detected emission
(including galaxies) is due to the intracluster light (ICL),
as from the statistical results obtained from 683 optically selected clusters
of galaxies at $0.2 \la z \la 0.3$ (Zibetti et al. 2005).

As a first main result, we find that the distribution
of the diffuse stellar emission around BCGs follows the distribution
of the intracluster medium (on a larger scale) in a system which is
either relaxed, like RXCJ\,2308.3$-$0211, or close to relaxation,
like RXCJ\,0232.2$-$4420.
In these two cases, the X-ray emitting, hot gas is expected to trace
the dark matter fairly well.

In these two X-ray luminous clusters, the $\mathrm{B-R}$ colour
of the diffuse light around BCGs is as red as the light
within the effective radii of the individual BCGs.
This holds whether there is a pair of BCGs (in RXCJ\,0232.2$-$4420)
or a single BCG (in RXCJ\,2308.3$-$0211) in the cluster core.
Hence, we confirm an overall consistency between colours of the ICL
and associated BCGs in relatively relaxed galaxy clusters
at intermediate redshifts (see Zibetti et al. 2005; Krick \& Bernstein 2007).
Conversely, we find that the $\mathrm{B-R}$ colour
of the diffuse stellar emission detected around the two pairs of BCGs
in the merging cluster RXCJ\,0014.3$-$3022
is significantly bluer than the central regions of the four BCGs.
If the contribution of intracluster stars to the diffuse stellar emission
around BCGs is not negligible, this means that the intracluster light
has multiple origins, possibly linked to the dynamical state of a cluster.

The $\mathrm{B-R}$ colours (observed frame) of the seven BCGs under study
and of the diffuse light around the three BCGs in the two X-ray luminous,
relatively relaxed clusters RXCJ\,0232.2$-$4420 and RXCJ\,2308.3$-$0211
can be reproduced by simple stellar populations modelled
as in Bruzual \& Charlot (2003) with at least solar metallicity
and age older than $2~\mathrm{Gyr}$.
The diffuse bluer stellar population around the four BCGs
in the X-ray luminous, merging cluster RXCJ\,0014.3$-$3022
must have a younger age and/or a lower metallicity
than the age and/or metallicity of the characteristic stellar population
of the BCGs.

The present picture of the ongoing large-scale merging in RXCJ\,0014.3$-$3022
favours an origin of the bluer diffuse light
detected around its overall ``red and dead'' BCGs
from shredding of star-forming, low-metallicty dwarf galaxies.

\begin{acknowledgements}
  Based on observations made with the ESO/MPG 2.2m telescope
  at the La Silla Observatory inside the Max Planck Gesellschaft (MPG) time.
  Based on observations made with the NASA/ESA Hubble Space Telescope,
  obtained from the data archive at the Space Telescope Science Institute.
  STScI is operated by the Association of Universities for Research
  in Astronomy, Inc. under NASA contract NAS 5-26555.
  D.P. thanks Ortwin Gerhard for a useful discussion regarding this work.
  D.P. acknowledges support by the German \emph{Deut\-sches Zen\-trum
  f\"ur Luft- und Raum\-fahrt, DLR\/} project number 50~OR~0405.
  F.B. acknowledges support by the International Max-Planck Research School
  (IMPRS) on Astrophysics.
  H.B. acknowledges support by
  \emph{The Cluster of Excellence ``Origin and Structure of the Universe''\/},
  funded by the Excellence Initiative of the Federal Government of Germany,
  \emph{EXC\/} project number 153.
\end{acknowledgements}

%__________________________________________________ Two column table
   \begin{table*}
      \caption[]{Basic properties of the sample clusters of galaxies.}
         \label{XrayProp}
     $$
         \begin{array}{p{0.20\linewidth}cccp{0.21\linewidth}}
            \hline
            \noalign{\smallskip}
            Denomination & {\mathrm{z}}^{\mathrm{a}} & {\mathrm{R}}_{500} / {[\mathrm{Mpc}]} & {\mathrm{L}}_{\mathrm{bol}} / {[\mathrm{10^{44}~erg\, s^{-1}}]} & Classification \\
            \noalign{\smallskip}
            \hline
            \noalign{\smallskip}
            RXCJ\,0014.3-3022 & 0.3066 & 1.24 & 21.2 \pm 1.7 & offset centre \\
            RXCJ\,0232.2-4420 & 0.2836 & 1.30 & 18.9 \pm 1.4 & primary with small secondary \\
            RXCJ\,2308.3-0211 & 0.2966 & 1.24 & 12.0 \pm 1.3 & single \\
            \noalign{\smallskip}
            \hline
         \end{array}
     $$ 
\begin{list}{}{}
\item[$^{\mathrm{a}}$] For the optical (i.e., spectroscopic) redshifts
see B\"ohringer et al. (2004).
\end{list}
\end{table*}
%-------------------------------------------------------------

%__________________________________________________ Two column table
   \begin{table*}
     \caption[]{Photometric properties of the brightest galaxies in the sample clusters.}
        \label{BCGProp}
     $$ 
        \begin{array}{p{0.19\linewidth}p{0.15\linewidth}p{0.16\linewidth}cccc}
           \hline
           \noalign{\smallskip}
           Denomination$^{\mathrm{a}}$ & RA(J2000.0) / [HMS] & Dec(J2000.0) / [DMS] & {\mathrm{r_{e,R}}}^{\mathrm{b}} / [{\mathrm{kpc}}] & {\mathrm{\mu_{e,R}}}^{\mathrm{b}} / [{\mathrm{mag~arcsec^{-2}}}] & {\mathrm{R_{tot,R}}}^{\mathrm{c}} / [{\mathrm{mag}}] & {\mathrm{(B-R)}}_{\mathrm{tot}}^{\mathrm{c}} / [{\mathrm{mag}}] \\
           \noalign{\smallskip}
           \hline
           \noalign{\smallskip}
           RXCJ\,0014.3-3022\,NW\,A & ~~~~~~~00:14:12.969 & ~~~~~~~-30:22:34.50 & 16.8 & 23.38 & 17.84 & 2.29 \\
           RXCJ\,0014.3-3022\,NW\,B & ~~~~~~~00:14:09.976 & ~~~~~~~-30:22:20.52 & 27.4 & 24.06 & 17.92 & 2.27 \\
           RXCJ\,0014.3-3022\,SE\,A$^{\mathrm{d}}$ & ~~~~~~~00:14:22.051 & ~~~~~~~-30:24:20.05 & 22.8 & 23.93 & 17.67 & 2.34 \\
           RXCJ\,0014.3-3022\,SE\,B$^{\mathrm{d}}$ & ~~~~~~~00:14:20.677 & ~~~~~~~-30:24:00.15 & 19.9 & 24.16 & 17.56 & 2.27 \\
           RXCJ\,0232.2-4420\,A     & ~~~~~~~02:32:18.522 & ~~~~~~~-44:20:47.81 & 17.6 & 23.25 & 17.31 & 2.17 \\
           RXCJ\,0232.2-4420\,B     & ~~~~~~~02:32:16.364 & ~~~~~~~-44:20:47.95 & ~~7.5 & 21.62 & 17.30 & 2.25 \\
           RXCJ\,2308.3-0211        & ~~~~~~~23:08:22.218 & ~~~~~~~-02:11:31.55 & 23.1 & 22.05 & 17.01 & 2.28 \\
           \noalign{\smallskip}
           \hline
        \end{array}
     $$
\begin{list}{}{}
\item[$^{\mathrm{a}}$] In a pair of BCGs, the letters ``A'' and ``B''
  identify the eastern and western components, respectively.
  No ranking in luminosity is used since total magnitudes are affected
  by deblending, the cluster cores being very crowded with galaxies
  (see Figs.~\ref{FigBR1a}--\ref{FigBR3}).
\item[$^{\mathrm{b}}$] R-band effective radii $r_{\mathrm{e,R}}$
  and surface brightnesses $\mu_{\mathrm{e,R}}$ refer to the inner
  ``de Vaucouleurs'' component of each BCG, except for the BCG
  of RXCJ\,2308.3$-$0211.
  For this object, effective parameters are derived
  from the empirical R-band surface brightness profile (see text).
\item[$^{\mathrm{c}}$] Total R-band magnitudes and $\mathrm{B-R}$ colours
  (observed frame) were determined after source photometry extraction
  with \emph{SExtractor\/} (Bertin \& Arnouts 1996) in a flexible
  (Kron-like, see Kron 1980) elliptical aperture with a Kron-factor of 2.5.
\item[$^{\mathrm{d}}$] The pair of BCGs RXCJ\,0014.3$-$3022\,SE
  is actually made of two very close pairs (see Fig.~\ref{FigXrayOpt1}).
  The total magnitudes and colours listed here
  belong to the two main components of this quadruple system
  (i.e., the BCGs RXCJ\,0014.3$-$3022\,SE\,A and B),
  as deblended through \emph{SExtractor\/}.
\end{list}
   \end{table*}
%-------------------------------------------------------------

%__________________________________________________ Two column table
   \begin{table*}
     \caption[]{Colours of the cluster brightest galaxies and the diffuse light around them.}
        \label{BCGICLCol}
     $$ 
        \begin{array}{p{0.19\linewidth}cccc}
           \hline
           \noalign{\smallskip}
           Denomination & {\mathrm{(B-R)_0}}^{\mathrm{a}} / [{\mathrm{mag}}] & {\mathrm{(B-R)}}({\mathrm{r_{e,R}}})^{\mathrm{b}} / [{\mathrm{mag}}] & \frac{{\mathrm{\Delta (B-R)}}}{{\mathrm{\Delta log~r}}}^{\mathrm{c}} / [{\mathrm{mag~dex^{-1}}}] & {\mathrm{(B-R)}}_{\mathrm{dsc}}^{\mathrm{d}} / [{\mathrm{mag}}] \\
           \noalign{\smallskip}
           \hline
           \noalign{\smallskip}
           RXCJ\,0014.3-3022\,NW\,A & 2.38 \pm 0.02 & 2.20 \pm 0.14 & -0.15 \pm 0.12 & 1.94 \pm 0.15 \\
           RXCJ\,0014.3-3022\,NW\,B & 2.38 \pm 0.02 & 2.09 \pm 0.16 & -0.21 \pm 0.12 & 1.94 \pm 0.15 \\
           RXCJ\,0014.3-3022\,SE\,A & 2.46 \pm 0.02 & 2.18 \pm 0.16 & -0.21 \pm 0.12 & 1.93 \pm 0.20 \\
           RXCJ\,0014.3-3022\,SE\,B & 2.42 \pm 0.04 & 2.26 \pm 0.11 & -0.13 \pm 0.09 & 1.93 \pm 0.20 \\
           RXCJ\,0232.2-4420\,A & 2.36 \pm 0.03 & 2.18 \pm 0.09 & -0.15 \pm 0.08 & 2.37 \pm 0.18 \\
           RXCJ\,0232.2-4420\,B & 2.29 \pm 0.01 & 2.27 \pm 0.06 & -0.02 \pm 0.07 & 2.37 \pm 0.18 \\
           RXCJ\,2308.3-0211 & 2.59 \pm 0.03 & 2.39 \pm 0.18 & -0.14 \pm 0.14 & 2.35 \pm 0.14 \\
           \noalign{\smallskip}
           \hline
        \end{array}
     $$
\begin{list}{}{}
\item[$^{\mathrm{a}}$] $\mathrm{B-R}$ colour measured
  in the BCG central region (i.e., at a radial distance of about 1 kpc
  along the major axis).
\item[$^{\mathrm{b}}$] $\mathrm{B-R}$ colour measured
  at the BCG effective radius.
\item[$^{\mathrm{c}}$] $\mathrm{B-R}$ colour gradient measured
  between the BCG effective radius and central region.
\item[$^{\mathrm{c}}$] $\mathrm{B-R}$ colour of the diffuse stellar component
 around each (pair of) BCG(s) - see text.
\end{list}
   \end{table*}
%-------------------------------------------------------------

%-------------------------------------------------------------
%-------------------------------------------------------------

%-------------------------------------------------------------
%-------------------------------------------------------------

%-------------------------------------------------------------
%-------------------------------------------------------------

%-------------------------------------------------------------
%-------------------------------------------------------------

%-------------------------------------------------------------
%-------------------------------------------------------------

%-------------------------------------------------------------
%-------------------------------------------------------------

%-------------------------------------------------------------
%-------------------------------------------------------------

%-------------------------------------------------------------
%-------------------------------------------------------------

%-------------------------------------------------------------
%-------------------------------------------------------------

%-------------------------------------------------------------
%-------------------------------------------------------------

%-------------------------------------------------------------
%-------------------------------------------------------------

\end{document}